\documentclass[preprint,authoryear,5p,times,twocolumn]{elsarticle}

\usepackage{graphicx}
\usepackage{amssymb}
\usepackage{amsmath}

\journal{New Astronomy Reviews}

\begin{document}

\begin{frontmatter}

\title{Super-massive binary black holes and emission lines in active galactic nuclei}

\author[aob]{Luka \v{C}. Popovi\'{c}}
\ead{lpopoovic@aob.rs}
\address[aob]{Astronomical Observatory, Group for Astrophysical Spectroscopy, Volgina 7, 11060 Belgrade, Serbia}

\begin{abstract}

It is now agreed that mergers play an essential role in the evolution of galaxies and therefore that mergers of supermassive black holes (SMBHs) must have been common. We see the consequences of past supermassive binary black holes (SMBs) in the light profiles of so-called `core ellipticals' and a small number of SMBs have been detected. However, the evolution of SMBs is poorly understood. Theory predicts that SMBs should spend a substantial amount of time orbiting at velocities of a few thousand kilometers per second. If the SMBs are surrounded by gas observational effects might be expected from accretion onto one or both of the SMBHs. This could result in a binary Active Galactic Nucleus (AGN) system. Like a single AGN, such a system would emit a broad band electromagnetic spectrum and broad and narrow emission lines.

The broad emission spectral lines emitted from AGNs are our main probe of the geometry and physics of the broad line region (BLR) close to the SMBH. There is a group of AGNs that emit very broad and complex line profiles, showing two displaced peaks, one blueshifted and one redshifted from the systemic velocity defined by the narrow lines, or a single such peak. It has been proposed that such line shapes could indicate an SMB system. We discuss here how the presence of an SMB will affect the BLRs of AGNs and what the observational consequences might be.

We review previous claims of SMBs based on broad line profiles and find that they may have non-SMB explanations as a consequence of a complex BLR structure. Because of these effects it is very hard to put limits on the number of SMBs from broad line profiles. It is still possible, however, that unusual broad line profiles in combination with other observational effects (line ratios, quasi-periodical oscillations, spectropolarimetry, etc.) could be used for SMBs detection.

Some narrow lines (e.g., [O\,III]) in some AGNs show a double-peaked profile. Such profiles can be caused by streams in the Narrow Line Region (NLR), but may also indicate the presence of a kilo-parsec scale mergers. A few objects indicated as double-peaked narrow line emitters are confirmed as kpc-scale margers, but double-peaked narrow line profiles are mostly caused by the complex NLR geometry.

We briefly discuss the expected line profile of broad Fe K$\alpha$ that probably originated in the accretion disk(s) around SMBs. This line may also be very complex and indicate the complex disk geometry or/and an SMB presence.

Finally we consider rare configurations where a SMB system might be gravitationally lensed by a foreground galaxy, and discuss the expected line profiles in these systems.

\end{abstract}


\end{frontmatter}

\section{Introduction}

The investigation of black holes is one of the most intriguing subjects in science and black holes have been a subject of investigation for over two centuries. In the last few decades investigations of black holes ranging in mass from sub-atomic masses (quantum black holes) to galactic masses (super massive black holes; SMBHs) have been carried out \citep[for more details see the review of][in this issue and references therein]{silk11}. We are concerned here with SMBHs. We would like to know how these enigmatic objects have grown over cosmic time to the sizes they have today. It is now recognized that massive galaxies grow through mergers and since it is well established that at least every massive galaxy harbors a supermassive black hole, one can expect the formation of super-massive binary black holes (SMBs) in the center of galaxies that have undergone major mergers \citep{b80,r81}.

The evolution of SMBs following the merger of their parent galaxies was first described by \cite{b80}, and the standard picture of SMB evolution is that the SMBs separation decreases, first through dynamical friction due to distant stellar encounters, then through gravitational slingshot interactions in which nearby stars are ejected at velocities comparable to the SMB orbital velocity when the orbital speed is comparable to the stellar velocity dispersion. Finally, the orbit decays through the emission of gravitational radiation, ultimately leading to coalescence of the binary \citep[see][]{mm05}. SMBs are expected to be among the primary sources of low-frequency gravitational waves detectable by the Laser Interferometer Space Antenna \citep[LISA, see e.g.][]{se05}. These waves will be useful to constraint cosmological parameters and to test general relativity

The gaseous accretion flow which forms around the SMBs (before the coalescence phase) can be a source of electromagnetic radiation \citep[see e.g.][]{mp05,kr10,ce10,st11}. The timescale, during which detectable electromagnetic radiation, rises to its maximum is governed by viscous diffusion of gas close to the remnant and ranges from several years to hundreds of decades in the case of SMBs.


On the other hand, the expulsion can take place under special conditions when two SMBs merge \citep[see e.g.][]{vo10}. The newly formed SMBH created after the merging process can then shoot out of the center of the system at high speed \citep[so called recoiling BH, see][]{ca07,gu11}. Over the last few years various predictions have been made about the speed at which the hole would be slung away \citep[see][]{go07,gu11}. The speed has been estimated to be from an order of 100s km s$^{-1}$ \citep[][]{do10} to several 1000s km s$^{-1}$ \citep[][]{ca07,lz11}. In principle, the speed of the SMBH mainly depends on the spin of the coalescencing black holes before merging. The case where the initial SMBH spins lie in the orbital plane is a configuration that leads to the so-called superkick and velocities can be significantly higher. For example, \cite{sp11} found several configurations where the kick-off velocity exceeds $\sim$ 15000 km s$^{-1}$. Such SMBHs would be recoiling in gas-rich galaxy merger remnants, and it has been claimed that one might expect some gas to accrete and  produce electromagnetic emission.

Putative SMB systems (or kicked-off SMBHs) have been used in explanation of a variety of observations, such as: a) stellar dynamical models of formation of cores in elliptical galaxies \citep[see e.g.][]{v89,mi02,me06,ks09}; b) the morphologies of some radio sources \citep[see e.g.][]{r93,ro00,ka10}; c) stellar/bulge structure and BH-bulge relations in galaxies \citep[see e.g.][]{se08,kb09,si11,b11}.

If an SMB is located in a nearly homogeneous gas cloud, different observational effect can be expected \citep[see, e.g.][]{bod10,f11}. The surrounding gas can accrete onto each of the SMB components \citep[see e.g.][]{do07,fa11}. In this case a binary Active Galactic Nucleus (AGN) system is present. Such system could potentially emit a broad-band electromagnetic spectrum from gamma to the radio band like a single SMBH. Each of the SMBHs could have a normal AGN structure that includes a narrow-line region (NLR), a dusty torus, an accretion flow emitting from the soft X-ray region to the IR, and a broad line region (BLR). Potentially emission from these regions could be used to detect a SMBs \citep{bod10,sc11}. As binaries enter their gravitational-wave dominated inspiral, they inevitably may induce large periodic shifts in the broad emission lines of any associated AGN \citep[see][]{lo10}.

 The suggestion of a connection between AGNs and SMBs is not new \citep[see][]{k68a,k68b,b80,r81,g83,g85,r85a,r85b,hf88}. The idea that AGN activity is triggered by mergers is even older.
In this scenario most of QSO host galaxies suffered mergers with accompanying starbursts that likely also triggered the QSO activity \citep[][]{benn08}, but the details of the physical mechanisms triggering AGN activity several hundred Myr after the merger remains a topic of active research. Moreover, there are a number of binary quasars at kpc-scale distance which have almost certainly arisen from mergers or interactions \citep[see e.g.][]{fvd09}.

The problem is that we do not observe pc and sub-pc scale SMBs. One can expect that SMB components have separations in the range 10$^{15}$-10$^{17}$ m \citep[e.g., 10$^{-4}$-10$^{-1}$ arcsec at 10 Mpc, see e.g.][]{b80}. It is hard to resolve a SMB with current optical telescope resolution. Direct imaging of binary AGNs on parsec scales can only be carried out in the local universe, and only in the radio band when both BHs are radio sources. Using radio interferometers (as e.g., the very long baseline array -- VLBA) one can resolve close binaries down to milli-arcsecond resolution. This method has been used to identify 0402+379 as an SMB candidate. It is a source with two compact radio cores seen at a projected separation of $\sim$7 pc, representing the closest system imaged to date \citep[][]{r06,ro09}. In the not too distant future, the {\it Gaia} mission should be able to resolve SMBs in the local universe \citep{p01,p11}, but the problem of resolving SMBs on higher redshifts will remain.

Since SMBs cannot be resolved in the optical, to detect a SMB, one should find some  effect showing that an SMB system's affect of its nearby environment. Examples included things such
as the BLR, or accretion disk \citep[see, e.g.][]{bo08,bo09a,bo09b}, and/or some manifestations of the motion in a two-body system, such as jet precession \citep{b80,r93,g96b,ro00,ka10},
double-peaked Balmer lines \citep{g96a,g96b}, high-shifted broad lines \citep[][]{bl09,er11,ts11} or quasi-periodic radio, optical, X-ray or $\gamma$-ray variation \citep[see e.g.][]{s88,v00,rm00}.


\begin{figure}
\centering
\includegraphics[width=3cm]{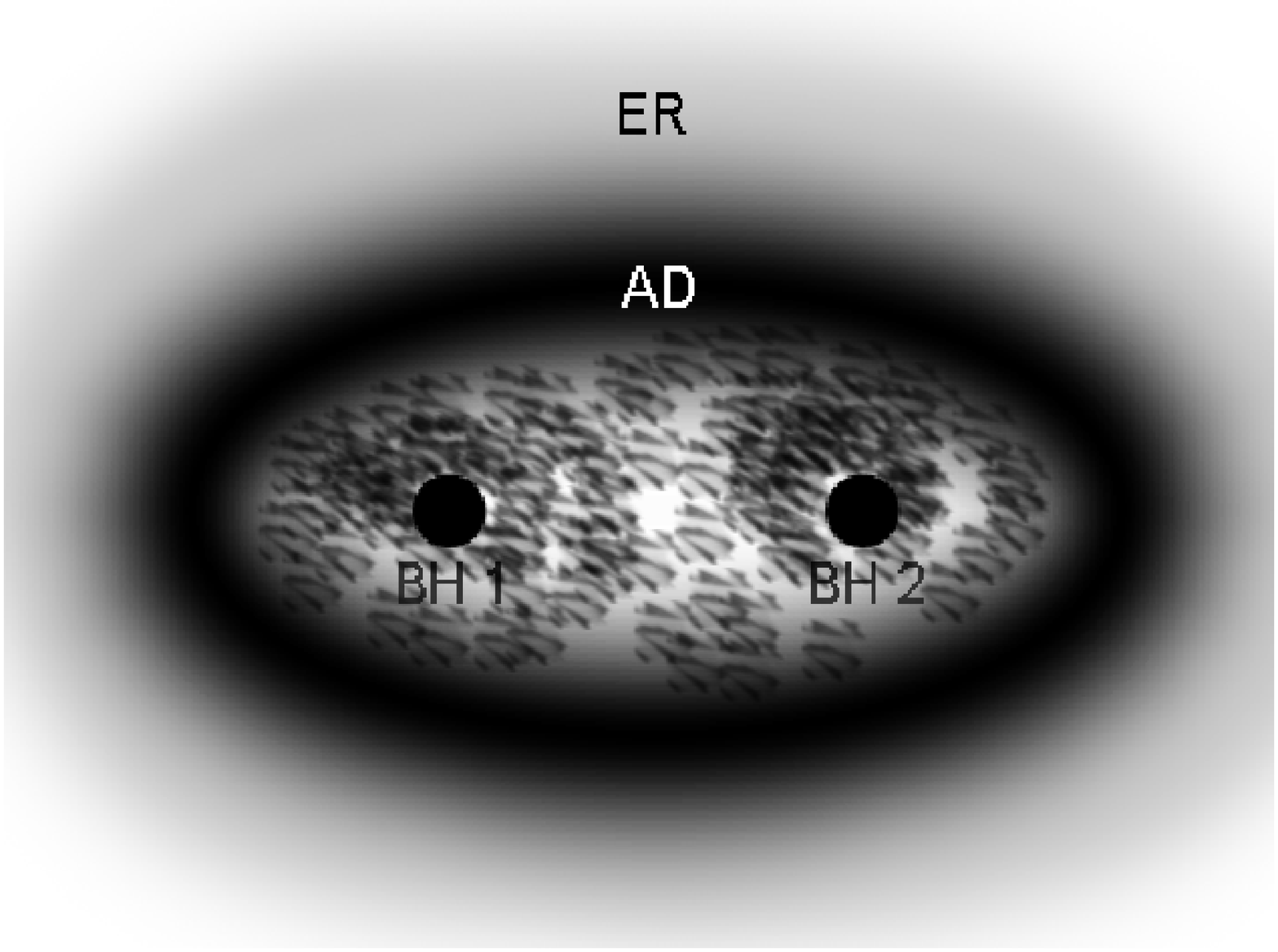}\\
\includegraphics[width=7cm]{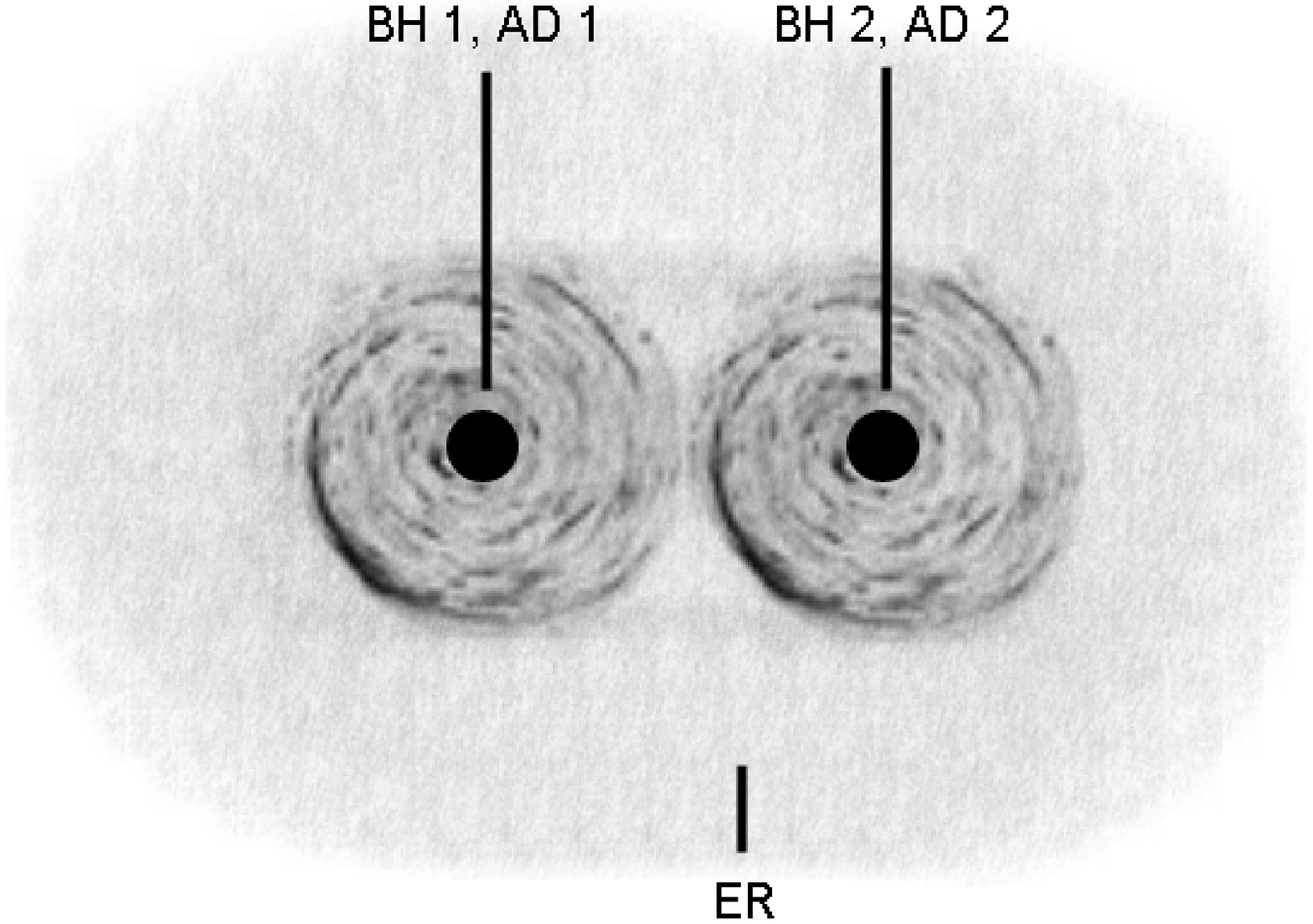}\\
\includegraphics[width=8cm]{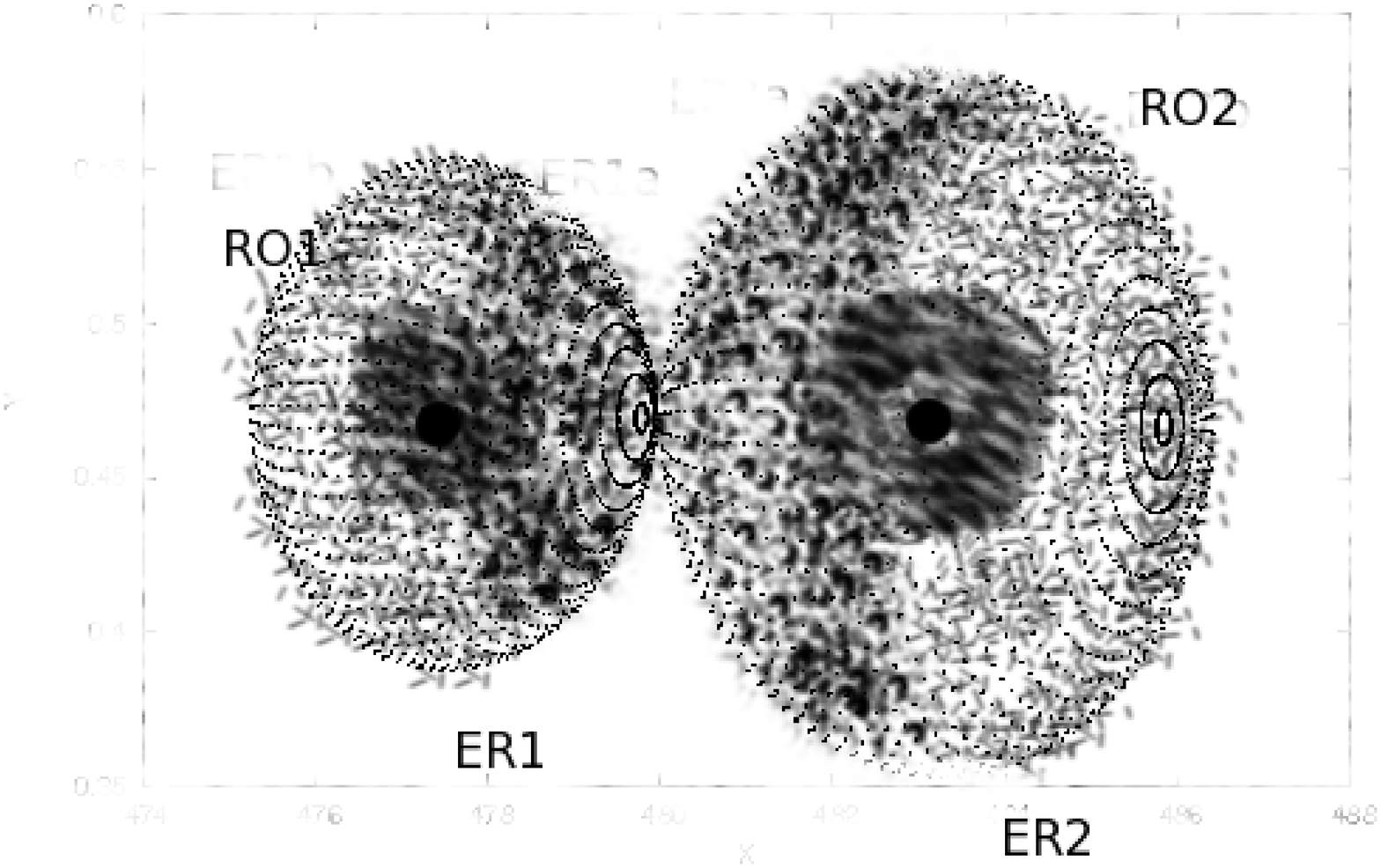}
\caption{The schemes of the possible SMB system: first  panel presents the SMB where only one accretion disk is present around two SMBHs that illuminates Emitting region (ER); second, each of SMBHs has an accretion disk and (semi-detached) emtting region(s) - ER;  and third is the scheme with two  clearly separated ERs, i.e. the Roche lobes (ROs) are separated.}
\label{f-bh}
\end{figure}

A number of galactic-scale binary AGNs has been discovered based on spatially resolved imaging or/and spectroscopy
\citep[see e.g.][]{k03,b08,c09,gr10,k11,li11}.
 The first discovered SMB system was, an ultra-luminous infrared galaxy, NGC 6240. The object was observed with the {\it Chandra} X-ray telescope that resolved
a pair of X-ray emitting AGNs \citep[][]{k03}. Two nuclei in NGC 6240 were revealed by the detection of absorbed, hard, luminous X-ray emission as well as two strong low-ionization Fe K$\alpha$ lines. The SMBHs in NGC 6240 have a projected separation of 1 kpc at a distance of 103 Mpc. {\it Chandra} detected a double with nucleus a 3.8 kpc separation (assuming a distance of 220 Mpc) in the luminous infrared galaxy Mrk 463 \citep[][]{b08}. \cite{gr10} found that SDSS J1254+0846 is a face-on, pre-coalescence merger consisting of two luminous radio-quiet quasars at $z = 0.44$, with a radial velocity difference of 215 km s$^{-1}$, separated by 21 kpc in a disturbed host galaxy merger with obvious tidal tails. \cite{k11} discovered a binary AGN in Mrk 739 with a 3.4 kpc separation at a distance of 130 Mpc, the binary AGN is particularly interesting because it shows no evidence of being an AGN in the optical, UV, or radio. Recently, \cite{li11} discovered  a kpc-scale, triple AGN, SDSSJ1027+1749, at z = 0.066. The system contains three emission-line nuclei, two of which are offset by 2.4 and 3.0 kpc in projected separation from the central nucleus. All three objects are classified as obscured AGNs based on optical diagnostic emission line ratios, with black hole mass  $M_{BH}\sim 10^8 M_\odot$ estimated from stellar velocity dispersions.

Spectroscopic studies may provide an alternative
method for revealing binary AGN, at even closer separations, through the search of velocity shifts in the multiple line systems resulting from the Keplerian motion of the two SMBHs.
However, it is interesting that of those AGNs, which have been detected as SMB candidates by spectroscopic methods \citep[see e.g.][]{g83,bl09}, all are
controversial because of other explanations for the same phenomenon or because of some inconsistency with other observational evidence \citep[]{er97,g10}.
 These binaries are at projected separations of the order less than a kpc, below which it is difficult to spatially resolve both SMBHs at cosmological distances.
Nevertheless, the emission lines of AGNs, at least, could indicate the possibility of SMB existence.

\cite{g83} suggested that some complex broad line profiles may be caused by SMBs in AGNs, i.e., that a quasar has two orbiting SMBs, each with an associated BLR. Nowadays, thanks to the databases containing a very large number of optical AGN spectra there has been a number of studies where both narrow and broad AGN emission lines have been used to identify possible SMBs detections \citep[see e.g.][etc.]{c09,bo09a,bo09b,bl09,x09,w09,sm10,li10a,li10b,c10,g10,k11,er11}.

X-ray continuum emission and the broad Fe K$\alpha$ line come from the innermost part of AGNs and some of the AGNs have been detected as binary systems using X-ray emission \citep[as e.g., NGC 6240, Arp 299, Mrk 463, see][]{k03,iv05,b08,wg09,wg10}.

In this paper we review and discuss the possibility of using emission lines to detect SMBs (or recoiling SMBH). We give more detailed discussion only of the possibility of a binary BLR detection using the broad optical lines, but we also give some brief discussion of the possibility of using iron K$\alpha$ lines and narrow lines as indicators of SMBs. In addition we consider the rare event of gravitational lensing of a binary AGN.

Paper is organized as following: In \S2 we discuss impact of the kinematics of a SMB system on the appearance of the broad-line spectrum considering different characteristics of
SMBs and different emission line regions. In \S3, \S4 and \S5 we consider the BLR, NLR, and Fe K$\alpha$ lines as indicators for detection of SMBs. In \S6 we discuss some unusual systems and in \S7 we outline our conclusions.

\section{Impact of the SMB kinematical parameters on BLR line parameters}

In considering lines as tools for finding SMBs, we should start from line parameters which can be readily measured, i.e., the line intensity, width, shift and shape. The line intensity depends mainly on the spectral energy distribution of the incident radiation and the physical properties of the emitting plasma.  One might therefore expect that special physical conditions in the plasma around a SMB might cause specific line intensity ratios emitted from the system \citep{mo11}.

On the other hand, widths, shifts and shapes of the BLR lines depend on the kinematics and geometry of emitting regions \citep{su00,g09}. The geometry and kinematics of the emitting region could well be different in an AGN binary system \citep[see e.g.][]{g83,bl09,er11,ts11,ba11}. Therefore, we need to consider the geometry and kinematics of the potentially complex emission region around the system as a whole and around each super-massive black hole. In this section we discuss some estimates of SMB kinematical parameters and their influence on line shifts, widths and shapes.  First we consider the expected line intensity difference between a single AGN and a close binary AGN.

\subsection{Line intensity as SMB presence indicator}

It is well known that line intensities depend on the plasma conditions \citep[density, temperature and relevant processes, see][]{of05}. In the case of  line-emitting regions of a single AGN the dominant energy source is the absorption of high-energy radiation from the inner regions of the accretion flow. In the case of a binary AGN system, there are two sources of ionization and two emission regions contributing to the composite spectrum. We can distinguish between two cases.  First there could be a common-envelope AGN system where the BLRs are semi-detached and there is a common envelope (see the first and second panels in Fig. \ref{f-bh}).
In this case there is mostly one emitting region around two SMBHs (with the one or two accretion disks as a source of ionization of the emitting region).  The second case is where the emitting regions are clearly separated around each BH (third panel in Fig. \ref{f-bh}). In these three specific cases, the illumination of gas in the BLR(s) would be different, consequently, one can expect that  ratios of lines emitted from ER(s) around SMBs may be different than ratios of lines emitted from a single SMBH.

So far only one study has considered the possibility that the line intensity ratio may be used as an indicator of an SMB's presence. \cite{mo11} investigated the spectral properties of the BLR in one SMBH that is a member of a sub-parsec SMB. They assumed that the binary, surrounded by a circumbinary torus, has cleared a gap, and that accretion occurs onto the secondary black hole fed by material closer to the inner edge of the torus. They considered C\,IV, Mg\,II and H$\beta$ line ratios as a function of orbital period, mass ratio and separation. They found that
the ratio $F_{Mg\,II}/F_{C\,IV}$ depends on the orbital separation (consequently of $P$) --- i.e., if the separation of the SMB is smaller, they obtained smaller
$F_{Mg\,II}/F_{C\,IV}$ ratios. On the other hand, the $F_{Mg\,II}/F_{H\beta}$ ratio may be significantly reduced only at the shortest separations.
Using a relatively simple model, \cite{mo11} were able to conclude that AGN spectra characterized by $F_{Mg\,II}/F_{C\,IV}<0.1$
could point out a signature of sub-parsec SMBs with the orbital periods between 0.1 -- 10 yr.

\begin{figure}
\centering
\includegraphics[width=5cm,angle=-90]{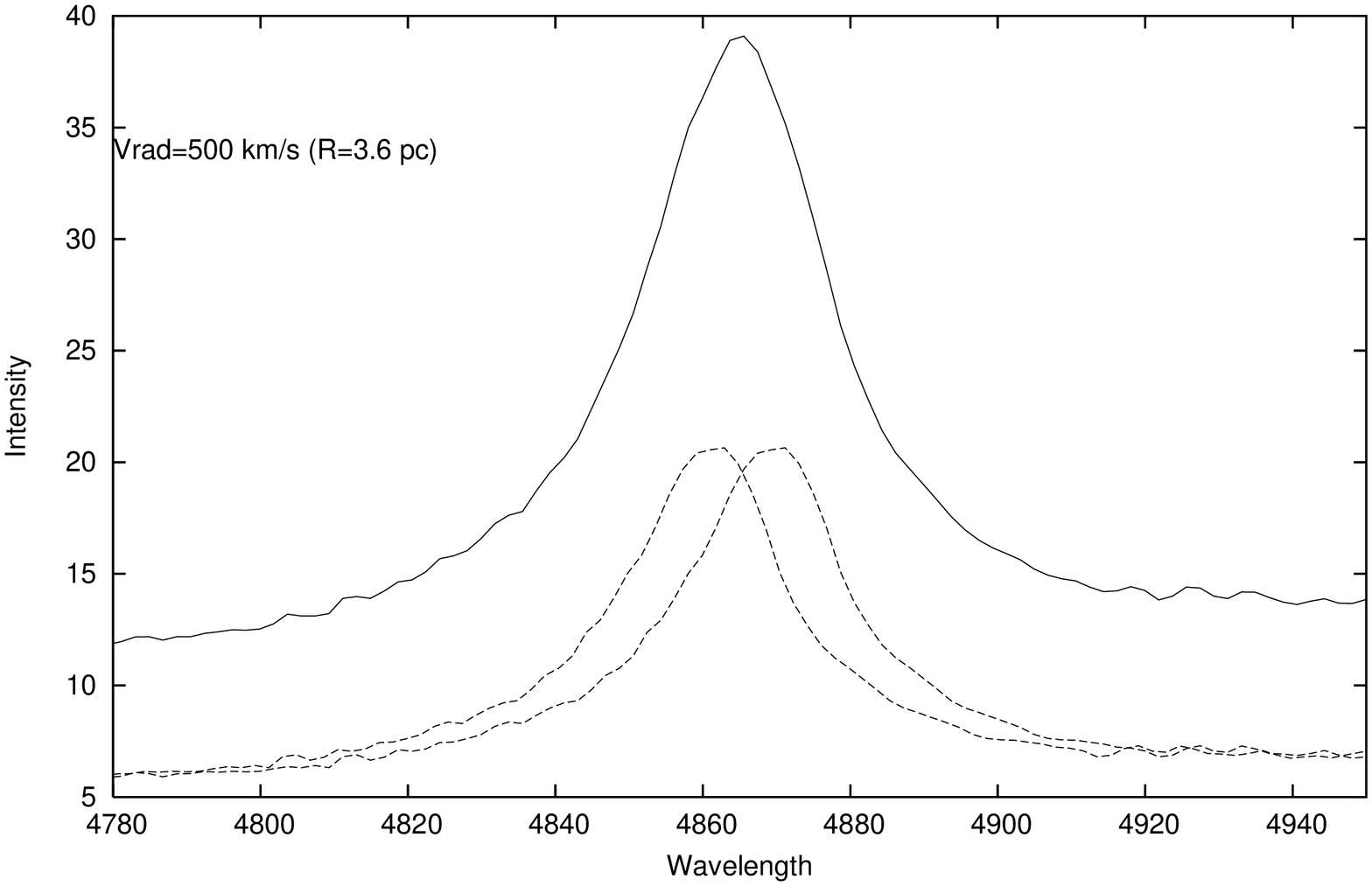}
\includegraphics[width=5cm,angle=-90]{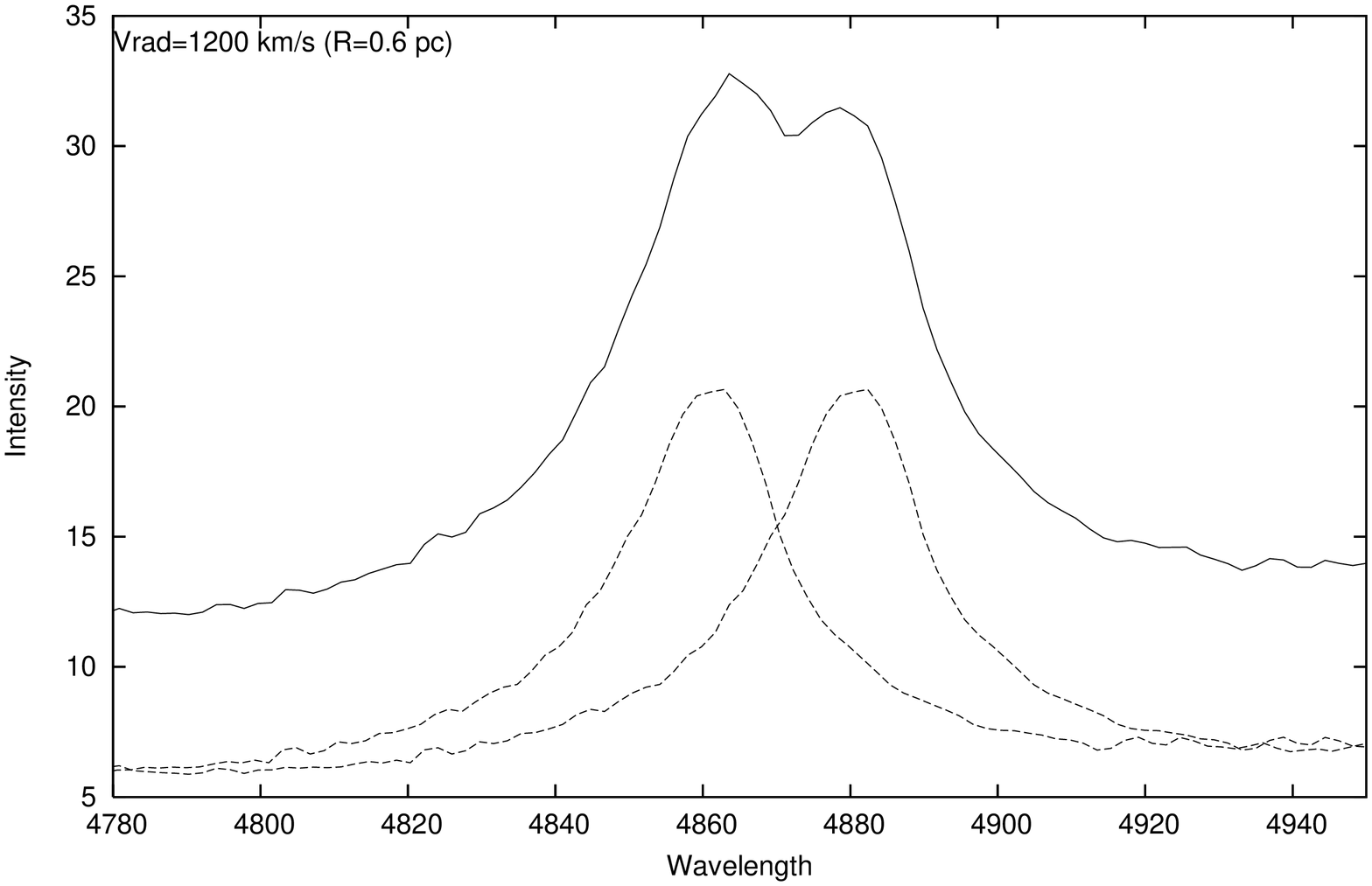}
\caption{Sum of the off-centered observed H$\beta$ line profile (AGN J00063+2012). First panel shows simulation where shift between component is 500 km s$^{-1}$ and second with shift of 1200 s$^{-1}$.}
\label{f-sp}
\end{figure}

\subsection{Expected kinematics in an SMB system: Influence on the line shift, width and shape}

The geometry of gas in close binary star systems has long been studied and it is instructive to first consider the SMB version of such a system.
Let us start with the simple approximation of taking an emitting cloud (EC) to be close to two massive objects and be gravitationally bounded, similar as in the case of the close massive binaries \citep[see e.g.][]{p71}.\footnote{This is the Roche lobe around a massive binary system when the gas is falling in through the inner Lagrangian point. It is approximately the  region bounded by a critical gravitational equipotential.}  We will assume a circular orbit of the BHs.  The geometry of the emitting gas should follow the shape of the equipotential surfaces \citep[according to the Roche model, see][and references therein]{p71,p00} that can be written as:

\begin{equation}
\Phi=\frac{R}{r_{1}}+\frac{R\cdot q}{r_{2}} +\frac{1+q}{2}[({x\over{R}}-\frac{q}{1+q})^2+{({y\over R})}^2]
-\frac{q^2}{2(1+q)}
\end{equation}
where $m_{1}$ and $m_{2}$ are the masses of the objects, $\Phi$ is the potential at an arbitrary EC, $R$ is distance between the black holes,
 $r_{1,2}$ the distances of the EC from the centers of gravity of the two
massive objects, and $x$ and $y$ are the EC Cartesian coordinates\footnote{The
origin of the rectangular system of Cartesian coordinates is at the
center of gravity of mass $M_{1}$, and the $x$-axis is the line joining the centers of the two massive object, and the
$y$-axis in the orbital plane}.
$q=\frac{m_{2}}{m_{1}}\leq1$ is the mass ratio\footnote{We will assume that $m_1$ is more massive BH (i.e., the ``primary'').}

As one can see from the Eq. (1), the most important parameter affecting the motion of the emitting clouds, and consequently geometry of the line emitting region in the SMB system is
the mass ratio.
There are two main parameters: the radial velocity of each component ($V_i$) and the orbital period ($P$) of the system.
$P$ governs line profile changes due to different velocities and position of the SMBHs, and it can be estimate by using the following relation
 \citep[see e.g.][]{p71,yl01}

\begin{equation}
P_{\rm orb} = 210\left(\frac{R}{0.1\rm pc}\right)^{3/2}
\left(\frac{2\times 10^8M_\odot}{m_1+m_2}\right)^{1/2} {\rm yr}
\end{equation}
Their radial velocities (in km s$^{-1}$) relative to the center of mass in the
line of sight can be estimated as \citep[][]{yl01}:
\begin{equation}
|v_i|=1.5\times10^3\left(\frac{0.1\rm pc}{R}\right)^{1/2}\left(\frac{m_1
+m_2}{2\times10^8M_\odot}\right)^{1/2}\left[\frac{2m_1m_2}{m_i(m_1+m_2)
}\right]\sin \theta_{\rm orb} \ \
\label{eq:vel}
\end{equation}
where $i$ denotes component with mass $m_1$ or $m_2$ $(i=1,2)$, respectively. $\theta_{\rm orb}$ is the orbital inclination to the line of sight.

The equations above can help us to roughly estimate and discuss kinematical parameters of a SMB system that may affect spectral lines emitted from the system.
From Eq. (1) one can roughly estimate
how much the SMB system affects the geometry (motion of ECs), and from Eq. (2) it is possible estimate the periodical (or quasi-periodicity) variations in the spectra and spectral lines. Eq. 3 shows how much the radial velocity can affect the composite line shapes, or how much a line can be shifted in the case of a single emitting region.

\begin{table*}
\begin{center}
\caption{Estimated $P$ and $V_{\rm rad}^{\rm max}$ for binary black holes with the same masses of $10^8\ M_\odot$ for characteristic distances. \label{t01}}

\begin{tabular}{|c|c|c|c|c|}

\hline
Distances& Corresponding line &FWHM in km s$^{-1}$ &$P$ in yr &$V_{\rm rad}^{\rm max}$ in km s$^{-1}$ \\
\hline
 {$10^{-3} - 10^{-2}$} pc & Fe K$\alpha$& $\sim 10^4-10^5$ &$\sim 2\cdot 10^{-4} - 7\cdot 10^{-3}$ &$1.5 \cdot 10^{4} - 5\cdot 10^3$ \\
 $10^{-2} - 1$ pc & Broad UV/optical lines&$\sim 10^3 - 10^4$&$\sim7\cdot 10^{-3} - 7\cdot 10^2$ &$\sim 5\cdot 10^3 - 5\cdot 10^2$ \\
 $10^{2} - 10^4$ pc & Narrow lines&$\sim \cdot 10^2 - 10^3$&$\sim \cdot 10^6 - 10^9$ &$\sim 50 - 0.5$ \\
 \hline\hline

\end{tabular}
\end{center}
\end{table*}

Considering Eqs. (1) one can conclude that there may be two cases: a) the case where $q<<1$ and b) $q\sim 1$ \citep[or let us say $q>0.01$, see
also][]{mo11}.

In the first case, it is not hard to see (from Eq. 1) that the motion of ECs will be affected only if the distances between the two BHs are small and only if the ECs are close to the both of the objects. SMBs with very low mass ratios (e.g., $q\leq 0.01$) cannot form in mergers of galaxies, since the dynamical friction time-scale is too long for the smaller SMBH to sink into the galactic center within a Hubble time \citep[][]{yu02}. Line shifts (see Eq. 3) can be seen only with the $m_2$ object, but there is major problem with whether such small mass object can still have an emitting line region.  In general the more massive BH picks up almost all gas infalling into the center. On the other hand, the emission that is coming from the smaller BH is unlikely to be intense enough to be comparable with emission of the regions around $m_1$.  It will therefore only make a very small (negligible) contribution to the emission of the more massive BH. We do not expect that SMBs with $q\leq 0.01$ will have a noticeable affect on line profiles.

In the second case, where the mass ratio is closer to unity, the geometry (or motion ECs) might be affected by the complex gravitational potential, and there can be various situations which
depend on the angle of view of an observer and the distances between super-massive black holes.

In the case $q\sim 1$ the distance between the BHs plays a significant role. There can be following situations: (i) the black holes are far away (on kpc scale for the NLR and pc, or even sub-pc scale for the BLR) so that each of them has an undisturbed AGN structure (i.e., each of SMBHs has the accretion disk and BLR/NLR), (ii) there are so called ``semi-detached'' systems (see Fig. \ref{f-bh})
\footnote{This is expected to be on a kpc scale  for the NLR gas and around a pc - scale for the BLR}  In this scenario the emission-line gas circulates from one BH to another, but the inner accretion disks (emitting in the soft X-ray region) are still un-disturbed. (iii) the distances are so small that there are binary SMBHs surrounded by a common envelope \citep[][]{b80,yl01,fa11}.
In the case (iii)
we might expect a high spectral (quasi-periodical) variability, especially the X-ray band \citep[see in more details][and reference therein]{mch10}.

It seems that two cases are perspective to be detected by AGN emission lines, the case i) and ii), but also it depends on the so called phase (the position of the two emitting regions with regard to the observer sight of view) of a binary system and inclination of the orbit. The influence of the phase on the line profile for case i) has been considered in a toy model given by \cite{p00} and depending of the mass ratio and the phase of a binary system the broad profiles can be different, from two-peaked (if there is no eclipsing between the objects) to the asymmetrical line profiles. This also depends on the velocity of the SMBHs and their periods \citep[see in more details][]{p00,s10}.
Note here that small eccentricity of the orbits could not significantly change the estimated parameters.

\subsubsection{Distances between BHs and stability of the line emitting regions}

Due to the gravitational interactions, an emission region may be destroyed or deformed, as, e.g., if the distance between SMBHs is of the order of magnitude or closer than dimensions of the emission region. In this case, the emission in lines can be complex, but here we will not consider this case. Here we will discuss only if line emitting regions are present in both or only in one SMBH (associated with one SMBH of SMBs or with recoiling SMBH). Let us first estimate the minimal distances between SMBHs for which the emission regions could still exist. Considering that gravitational interaction decreases with distances as $\sim r^{-2}$, there is necessarily that distances between SMBHs are significantly larger than sizes of the emitting regions. Consequently, the distances between the SMBHs in the system should be at least two order of magnitude larger than the size of the emission region.

To give some estimates of the characteristic distances, we took into account the estimated sizes of the line emitting regions in AGNs: the Fe K$\alpha$ line is coming from the innermost region (several to several 100s gravitational radii), and dimensions are of the order of $10^{-5} - 10^{-4}$ pc \citep{n07}. The broad line region is more extensive, at least an order of magnitude and we assumed dimensions of $10^{-4} - 10^{-2}$ pc \citep{ka00,pe04} and at the end, the NLR can be significantly extensive \citep[see e.g.][]{be06,be06b,be06c} from several pc (we took 10 pc) to several kpc (we took 1kpc). Taking different scales, one can expect that different emission region will be affected in SMBs and consequently different effects will be present in the line profiles.

Beside mentioned emitting regions, according to the standard unification model of AGN, it is expected that the material feeding the AGN (in the case of a single SMBH) is in the form of a torus \citep[see][]{an93}. Expected dimensions of the torus are several pc \citep[see e.g.][]{sta11}, therefore,   in the sub-pc scale  SMBs one cannot expect a classical torus. However, the surrounding gas formed in the interaction of SMBHs can play a role (similar as a torus) in feeding the AGN, i.e. provide material for accrete onto each of the SMB components \citep[see e.g.][]{do07,fa11}.

In Table \ref{t01} we give three characteristic interval of distances between SMBHs with respect to the dimension of the Fe K$\alpha$, Broad and Narrow line emitting regions. Also in the Table \ref{t01} we give estimates for the maximal radial velocities of components and their orbiting periods for characteristic distances. The estimates are given for the SMB system with the same mass of $10^8\ M_\odot$.

It is interesting to compare the typical FWHM of lines from three emitting regions with the corresponding maximal radial velocity. As it can be seen, the maximal radial velocities (orbits are edge-on) are in principle smaller than FWHM. It seems that is very unlikely to see clearly split lines, rather one can expect asymmetry or peaks in the line profile. As an illustration in Fig. \ref{f-sp} we simulate the composite broad H$\beta$ line (one AGN J00063+2012), taking that the same spectra are coming from a binary BLR with velocities of 250 km s$^{-1}$ each (or estimated distance between them around 3.6 pc if orbit is edge-on oriented) and 600 km s$^{-1}$ (distance between components is around 0.625 pc for the edge-on oriented orbit); as it can be seen from the Figure, in the first case we cannot see any influence of SMBs on line shapes, and in the second case, only two peaks can be registered in the line profile.

On the other hand, due to periodical orbiting of the binary, the periodical (or rather quasi-periodical) spectral oscillations may be present in the Fe K$\alpha$ line (from several hours to several days) and also in the broad lines (order from several days to several hundred of years). Moreover, the gas can periodically transfer from the circumbinary disk to the black holes when the binary is on an eccentric orbit and produce periodical oscillations in the electromagnetic spectra \citep[see][]{hay08,cu09,gil10}. Consequently, variability in the line profile due to different position and radial velocity of the SMB components, and periodical transfer of gas is expected to be observed.

\section{Broad emission lines and SMBs}

\begin{figure}
\centering
\includegraphics[width=8cm]{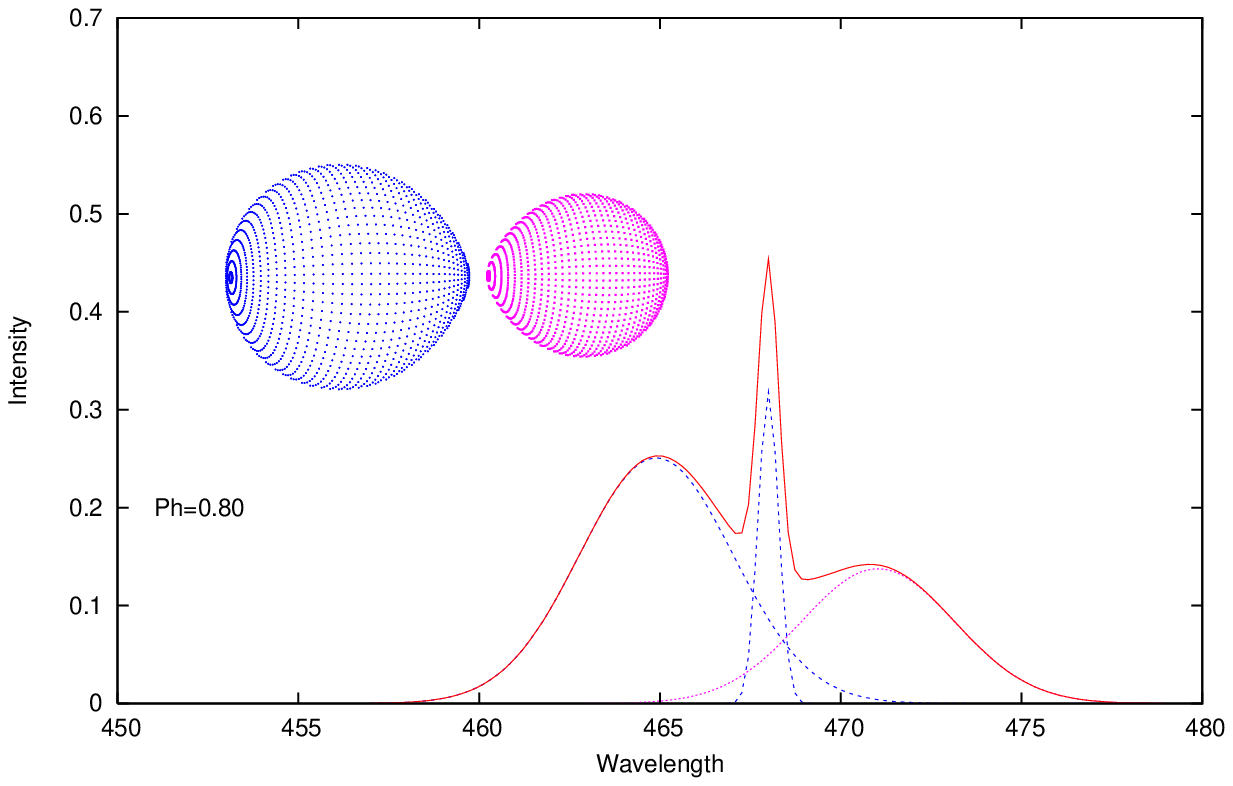}
\includegraphics[width=8cm]{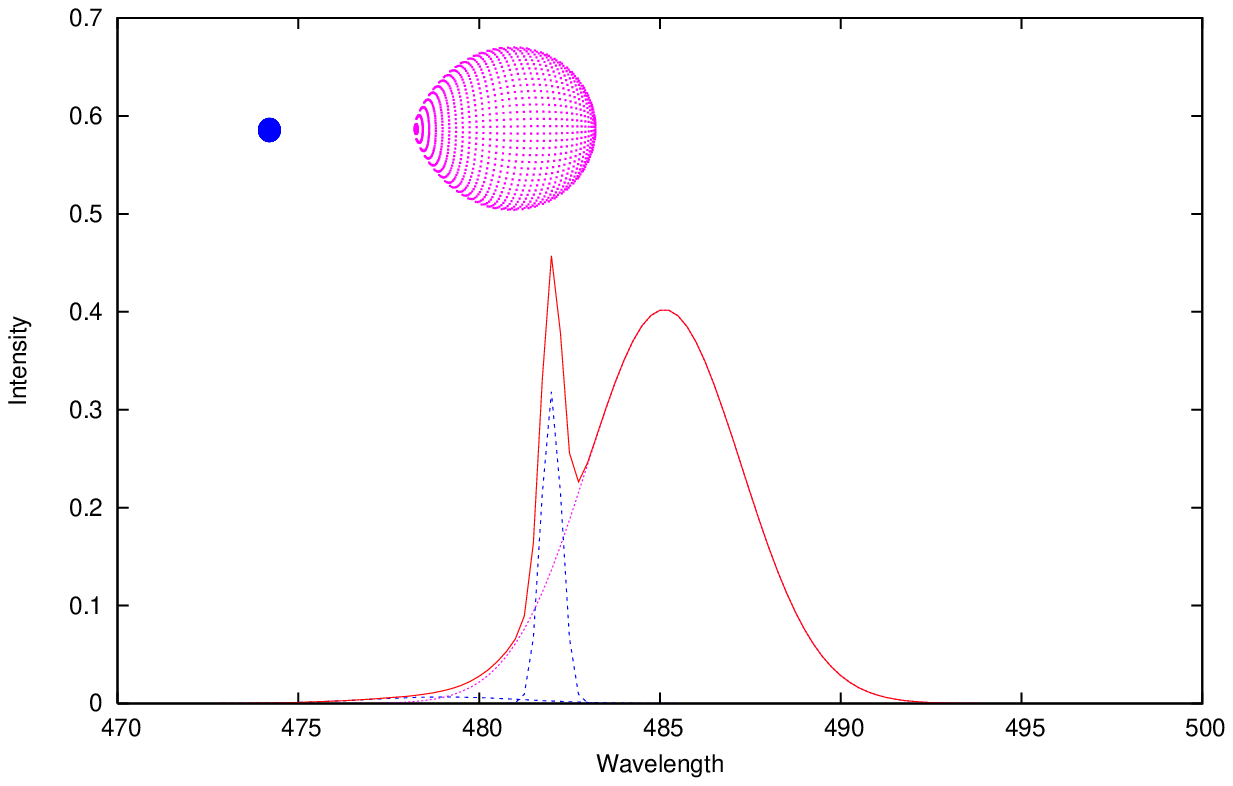}
\caption{The H$\beta$ line profile emitted from the SMB system for the cases: 1) both of black holes have the BLR (up) and only one component has the BLR (down). The scheme of SMB system is present on the plot. Simulation of the binary broad line region is performed as described in \cite{p00}.}
\label{f001}
\end{figure}

Considering a SMB system, one should take into account possibility that
one (or both of the two SMBHs) in the system has a gaseous BLR that emits the broad, permitted
lines, typical for quasar spectra. The orbital motion of the SMB causes the lines to shift periodically making the spectrum analogous to that of
a single- or double-lined spectroscopic binary star \citep[][]{k68a,k68b,b80}. \cite{g83,g85} suggested that close supermassive
binary black holes could be detected through their effect
on the broad line profiles. In the first instance, displaced broad
line peaks could be a consequence of the orbital motions
of close binary black holes (see Figs. \ref{f001} and \ref{f002}).

One of the first candidate for the SMB system was 3C390.3 \citep[][]{g83,g85}. Moreover, an observed systematic drift of the blue-shifted, broad
H$\beta$ peak in the spectrum of 3C 390.3 between 1968 and 1988 was consistent with the SMB hypothesis \citep{g96b}. But
reverberation mapping of 3C 390.3, however, showed that on a light-crossing timescale the red and blue peaks varied near simultaneously \citep[see e.g.][]{die98,sh01,sh10} and the other effects seen in the line profiles indicated disc origin \citep[see][]{er97,sh01,eh03,sh10,pop11}. This
strongly rules out binary black holes as the cause of
the displaced peaks in this object. It is likely that disk emission is present in the double peaked (and strongly asymmetric) emitters \citep[see][]{st03}. Nevertheless, the binary
black hole hypothesis remains a reasonable explanation
for broad, single-peaked (and also double-peaked) Balmer lines that are shifted
from their nominal wavelengths \citep[][]{er11,ts11}.

In the case of the SMBs, peculiar broad spectral lines with large velocity shifts of the order of $\sim$ 10$^3$ kms$^{-1}$ are expected (see Table 1). Recently, AGNs with such large velocity shifts are selected among thousands
of quasar spectra, and several spectroscopic SMB candidates have been found in archive
of Sloun Digital Sky Survey (SDSS), as e.g.: J092712.65+294344.0 \citep[][]{kz08,bo09a,sh09a,do09b}, J153636.22+044127.0 \citep{bl09,tg09}, J105041.35+345631.3 \citep{sh09}, J100021.80+223318.6, \citep[4C+22.25, see][]{de10a}, E1821+643 \citep{rob10}, J093201.60+031858.7 \citep[][]{ba11}, 12 candidates have been presented in \cite{er11} and nine in \cite{ts11}.
All these sources show shifted line systems, but their spectra
differ significantly one from another and the interpretation
of their physical nature is not unique \citep[see][]{g10,er11,ts11}.

To illustrate expected line profiles, we modeled the binary BLRs around two SMBs in the same way as it was
described in \cite{p00}. Close to each black hole the structure of the BLR can be similar to the structure inferred for BLRs around single SMBHs.  We simulate the H$\beta$ line profile for a system with following parameters: mass ratio of $q=0.5$, distances between BHs $R=0.3$ pc, orbiting period of $P=600$ years. First we illustrate in Fig. \ref{f001} (panel up) the case where both SMBHs have the BLR (as the Roche lobe around each of them)\footnote{ Note here that in a SMB the sizes of the BLRs are limited to be smaller than the Roche lobes, but it is not important for these simulations, since here we will only simulate the line profiles from a SMB system with two BLR and this approximation will not affect the final result}, where the orbit is edge-on to the observer. Also, we arbitrarily added a narrow component (narrow central Gaussian). As it can be seen in Fig. \ref{f001}, in the case of binary BLR, the line profile is composite from emission of both regions, and line profiles (see Fig. \ref{f002}) can be quite different depending on the phase \citep[see][]{p00,s10}. The line profile may change from a very asymmetric (double peaked and highly shifted) profile to the symmetric profile. It is characteristic that line profiles change the shape, and variability in the line profile can be expected.

As it is shown in Figs. \ref{f001} and \ref{f002} in some cases the line
profile can show two peaks in the case of a binary BLR system, but taking into account the estimated maximal radial velocities
(see Table \ref{t01}) for characteristic distances between SMBs, one can expect an asymmetric and off-centered line shapes rather than two-peaked ones.

\subsection{Two or one broad line emitting region in SMBs?}

In a SMB system, there is a possibility that both SMBHs have emitting regions, or that only one SMBH has an emitting region.

\subsubsection{Binary BLR}

In the evolution scenario of an unequal mass binary black hole with a coplanar gas disk \citep[see][]{es04,do07} that contains a gap due to the presence of the secondary black hole \citep[early work of][for stellar binary-disk interaction]{al94} is expected. It means that the secondary has chance to have a BLR, since viscous evolution of the outer circumbinary disk initially hardens the secondary. On the other hand, the inner disk drains on to the primary (central) black hole. Therefore, the gaseous mass profile bound to each black hole that can lead to the formation of two small accretion disks \citep[][]{do07}. Discs will illuminate the gas around and, consequently, two BLR may be formed and line profiles emitted from the SMB system may be composed from the two rotating BLRs. Additionally, a high eccentricity in a SMB orbit will led that emerges from migration phase will contribute to the possibility off triggering of periodical inflows of gas onto both SMBHs \citep{rod11}, that can cause variability (periodical or quasiperiodical) in the line profiles.

\begin{figure*}
\centering
\includegraphics[width=3.6cm]{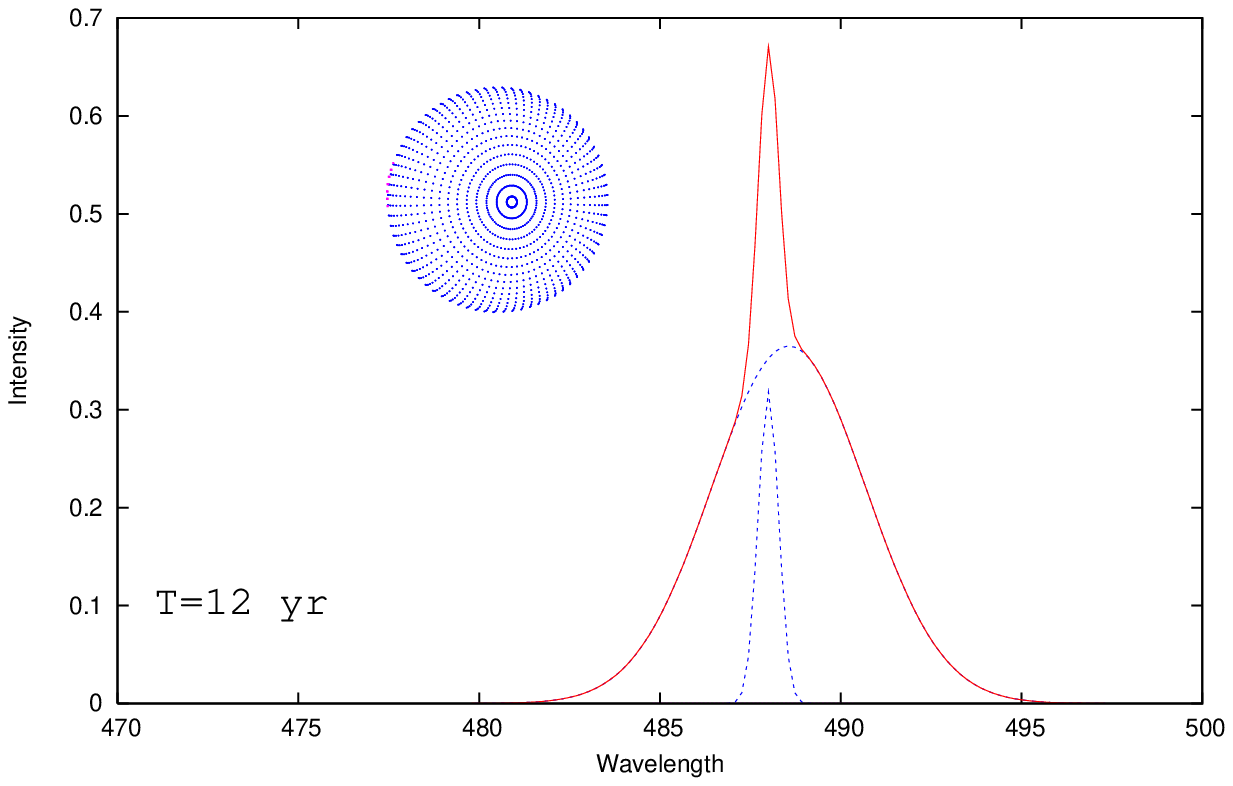}
\includegraphics[width=3.6cm]{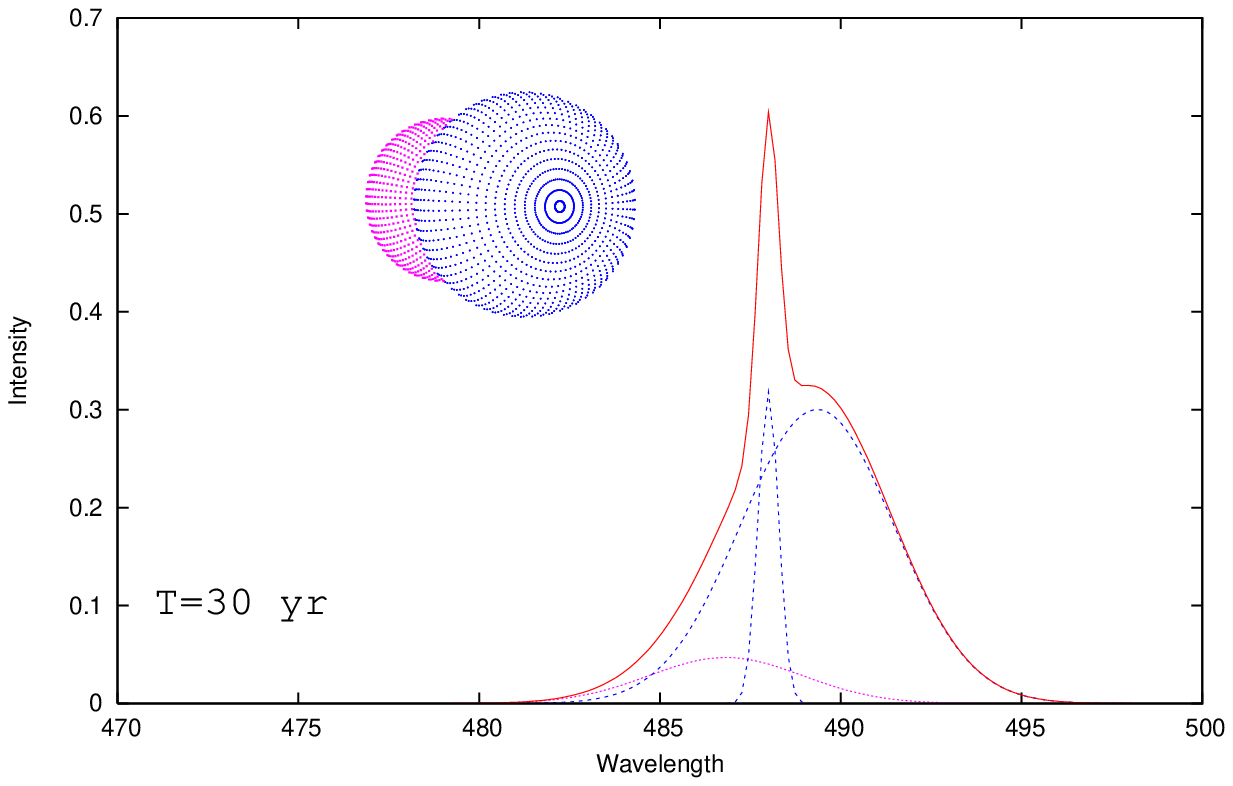}
\includegraphics[width=3.6cm]{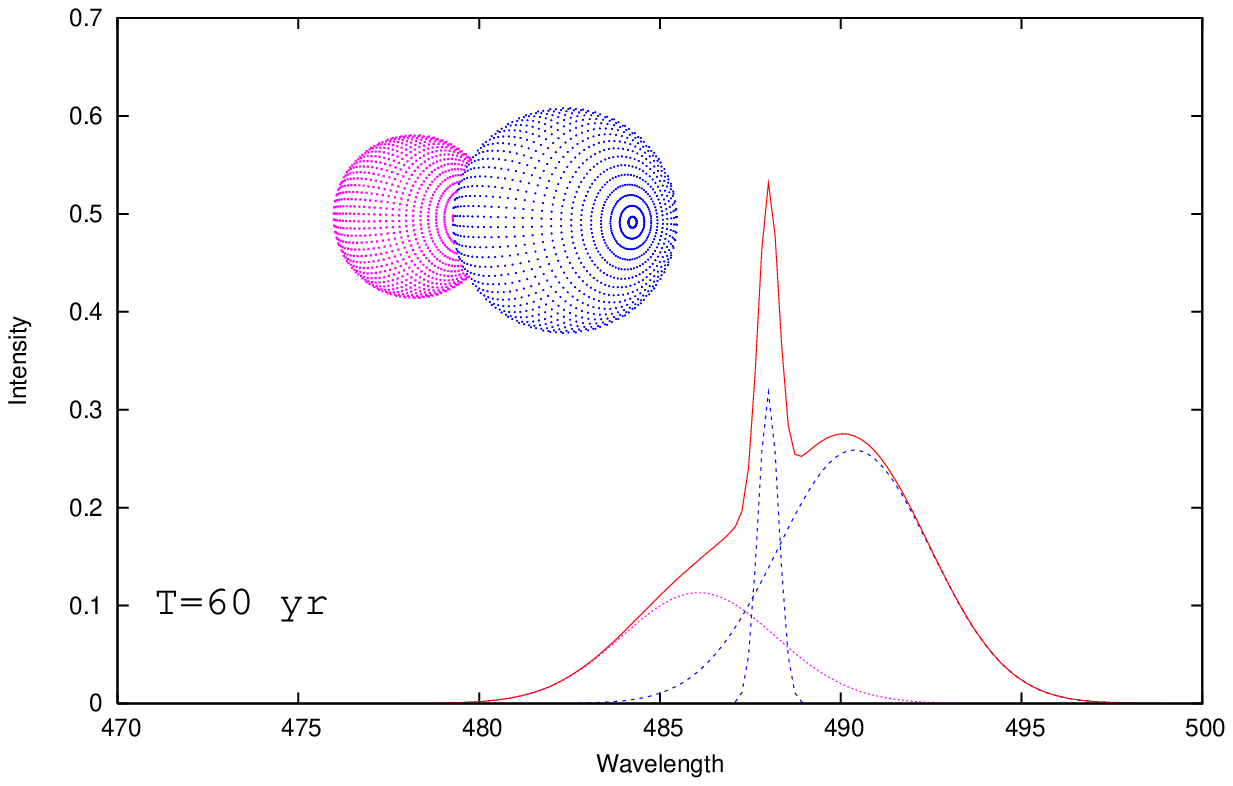}
\includegraphics[width=3.6cm]{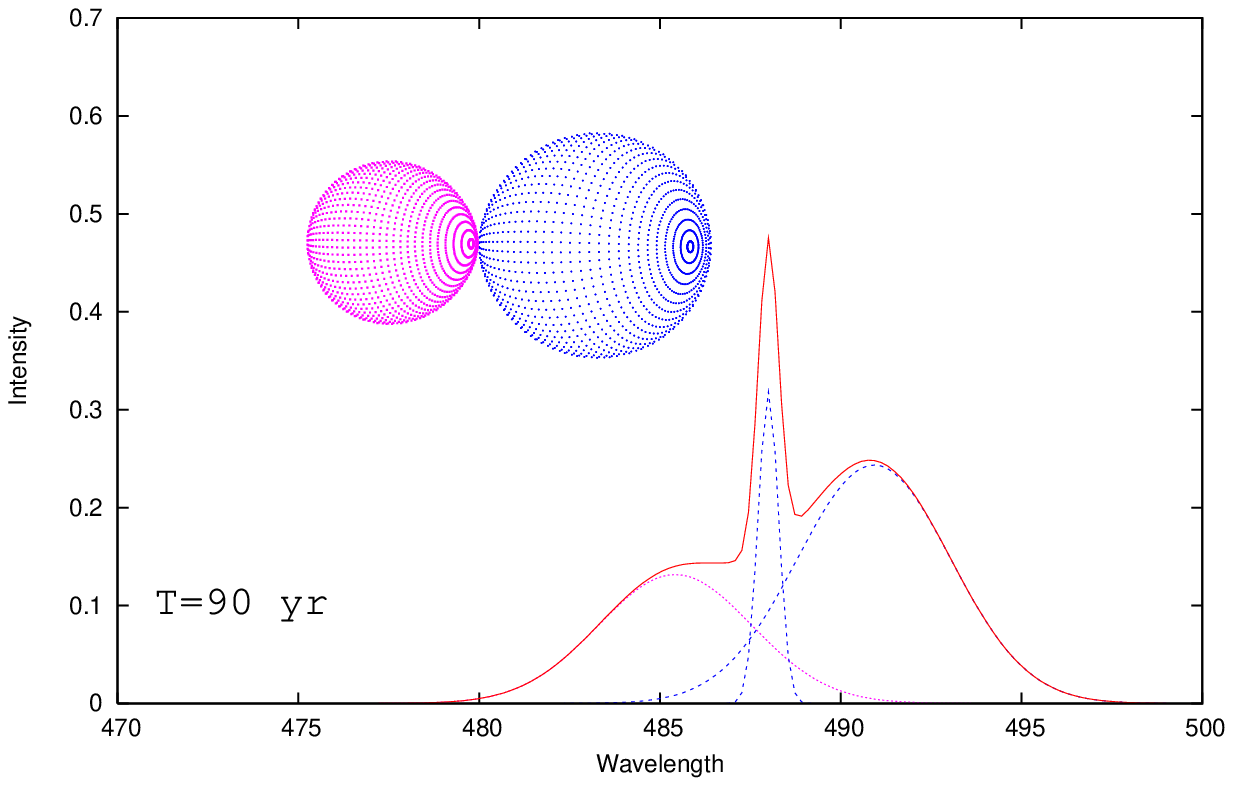}
\includegraphics[width=3.6cm]{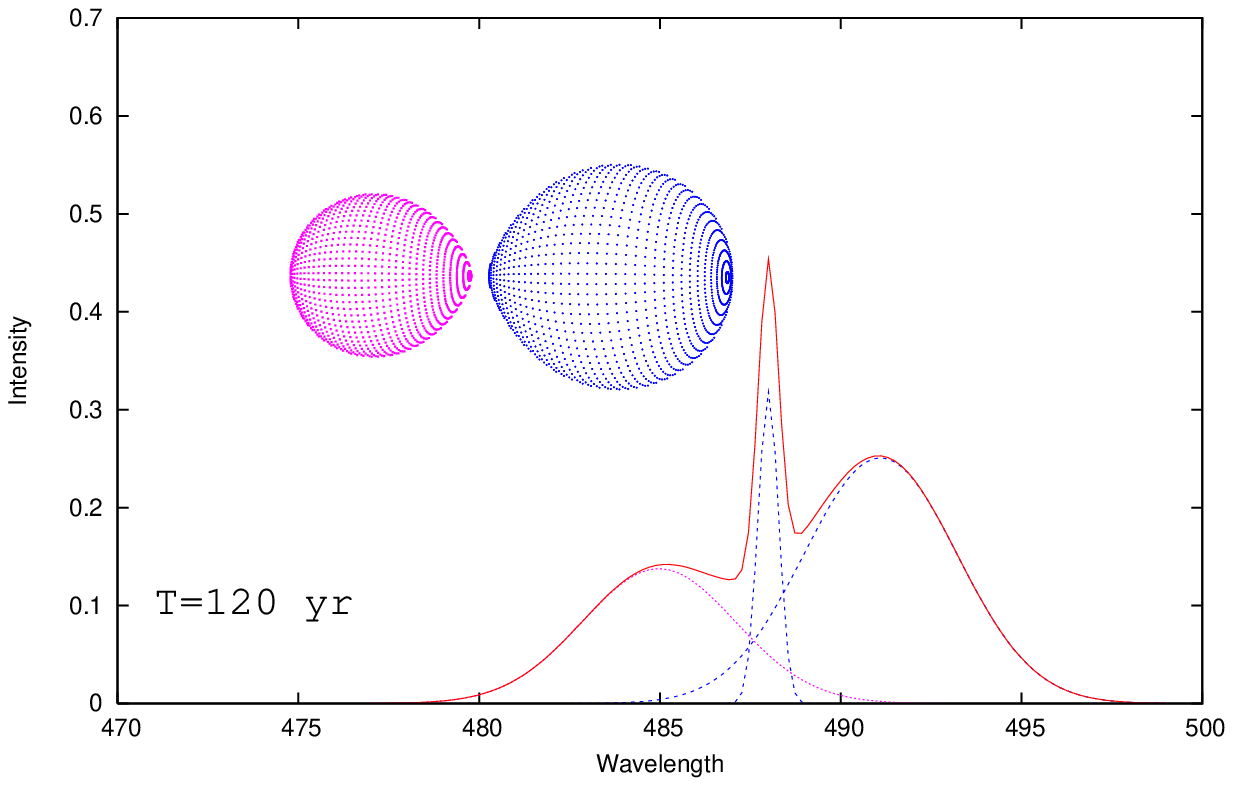}
\includegraphics[width=3.6cm]{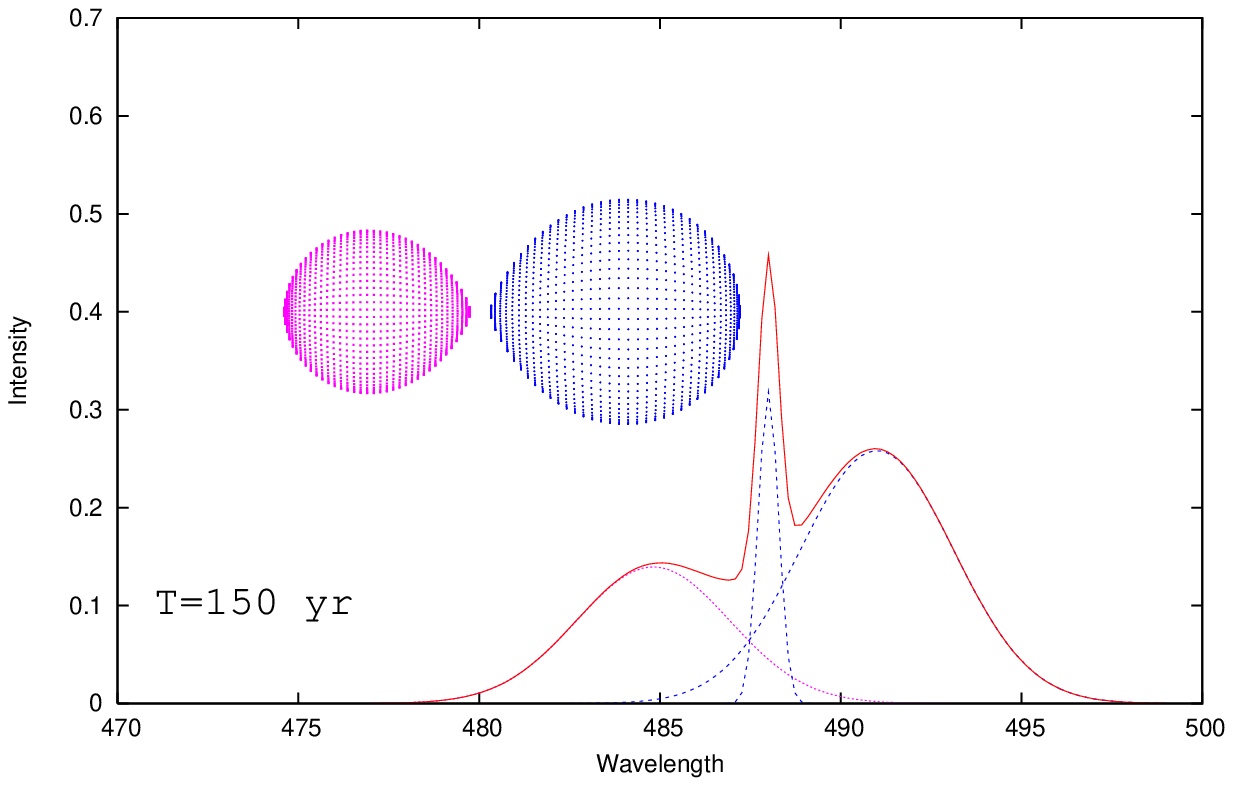}
\includegraphics[width=3.6cm]{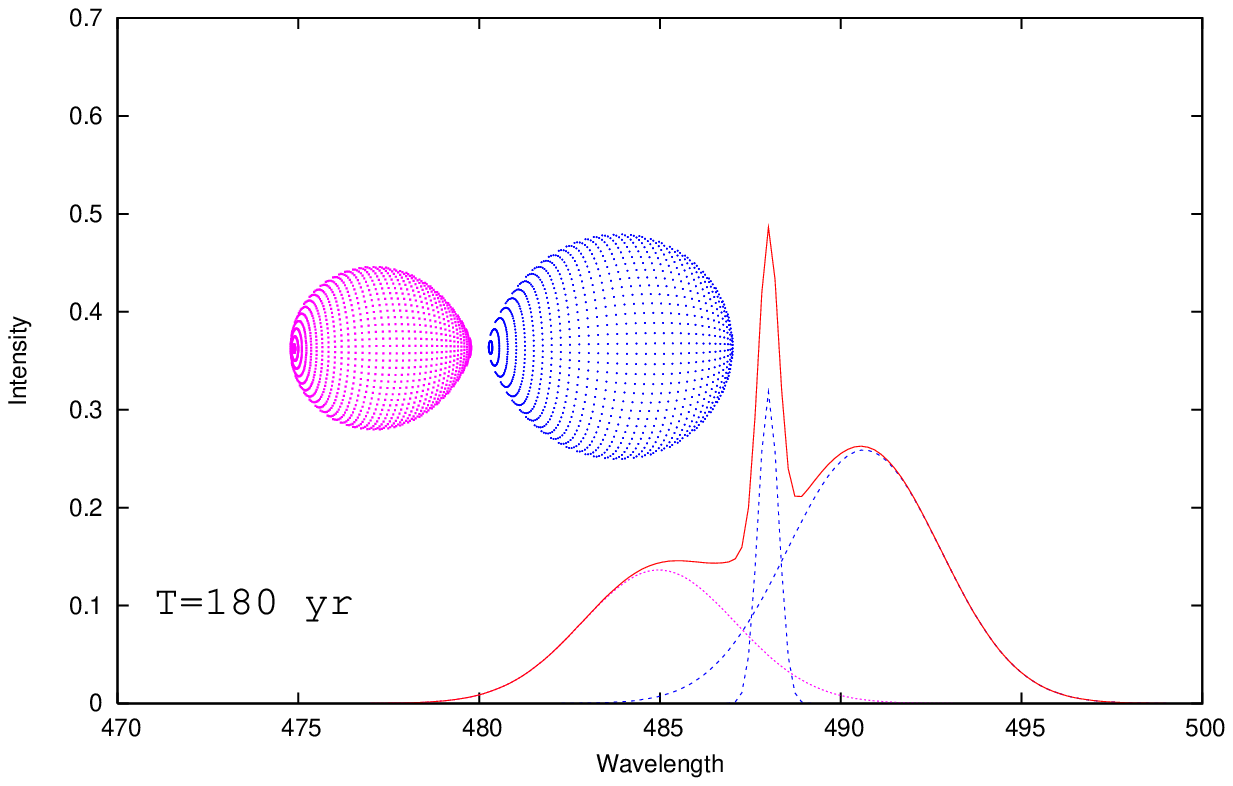}
\includegraphics[width=3.6cm]{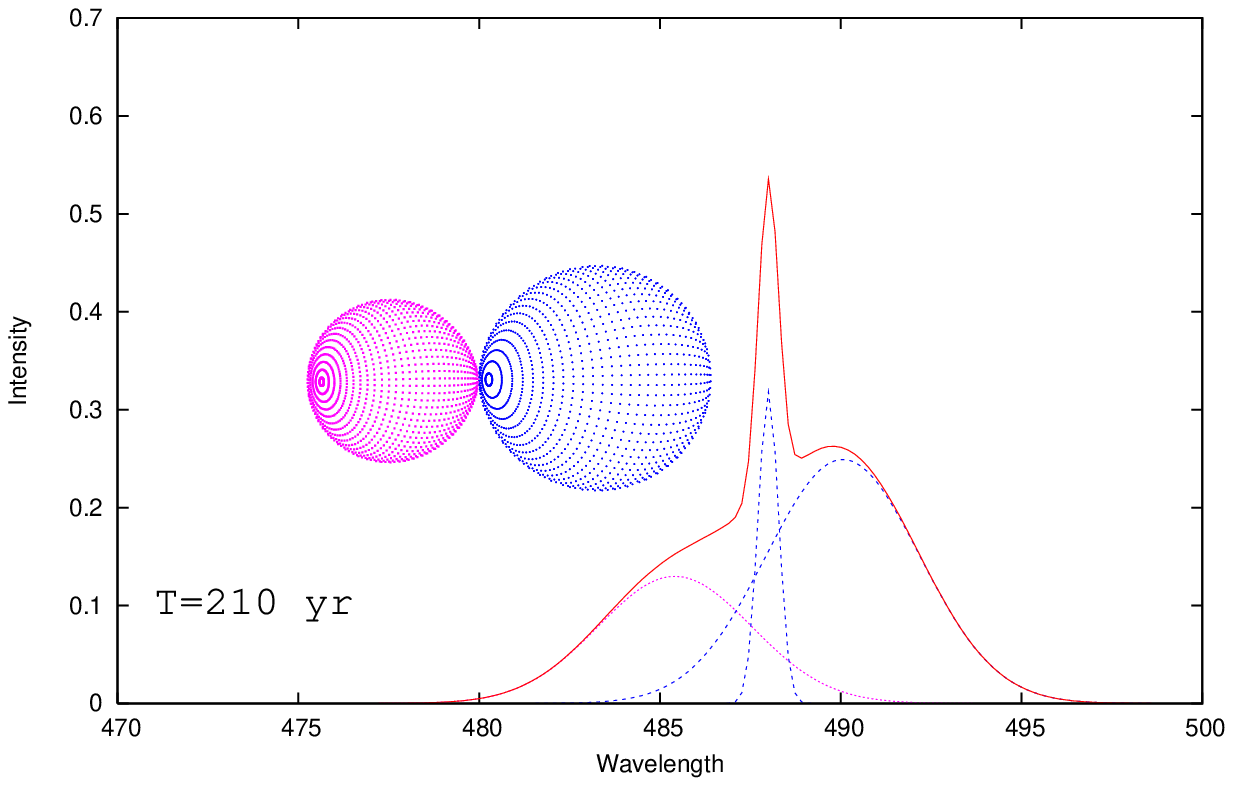}
\includegraphics[width=3.6cm]{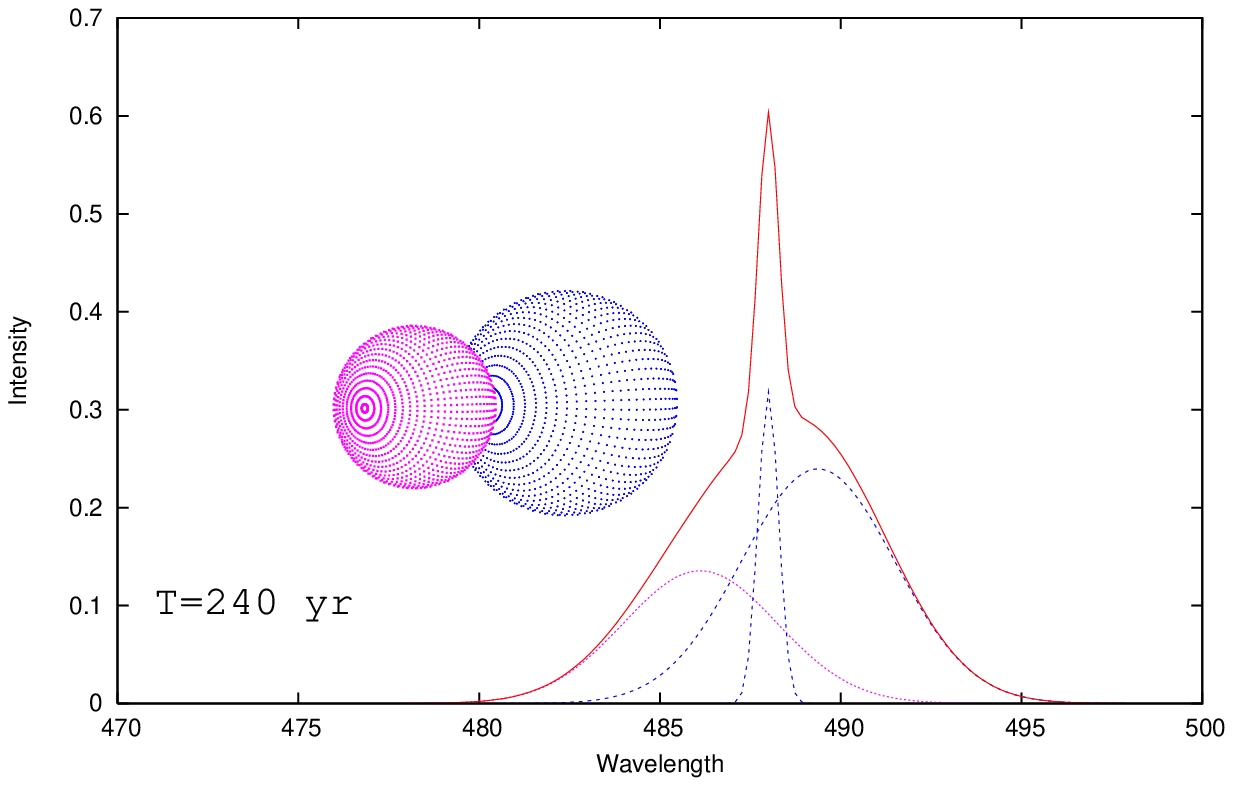}
\includegraphics[width=3.6cm]{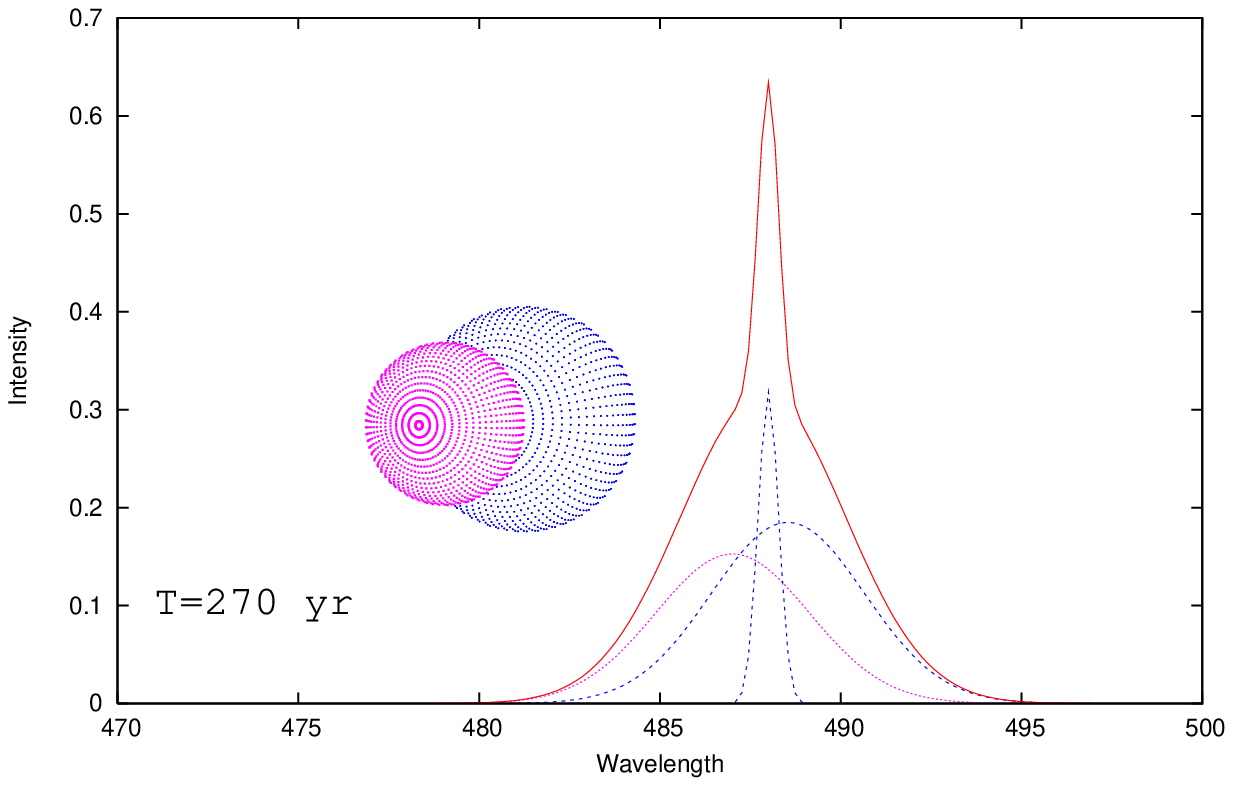}
\includegraphics[width=3.6cm]{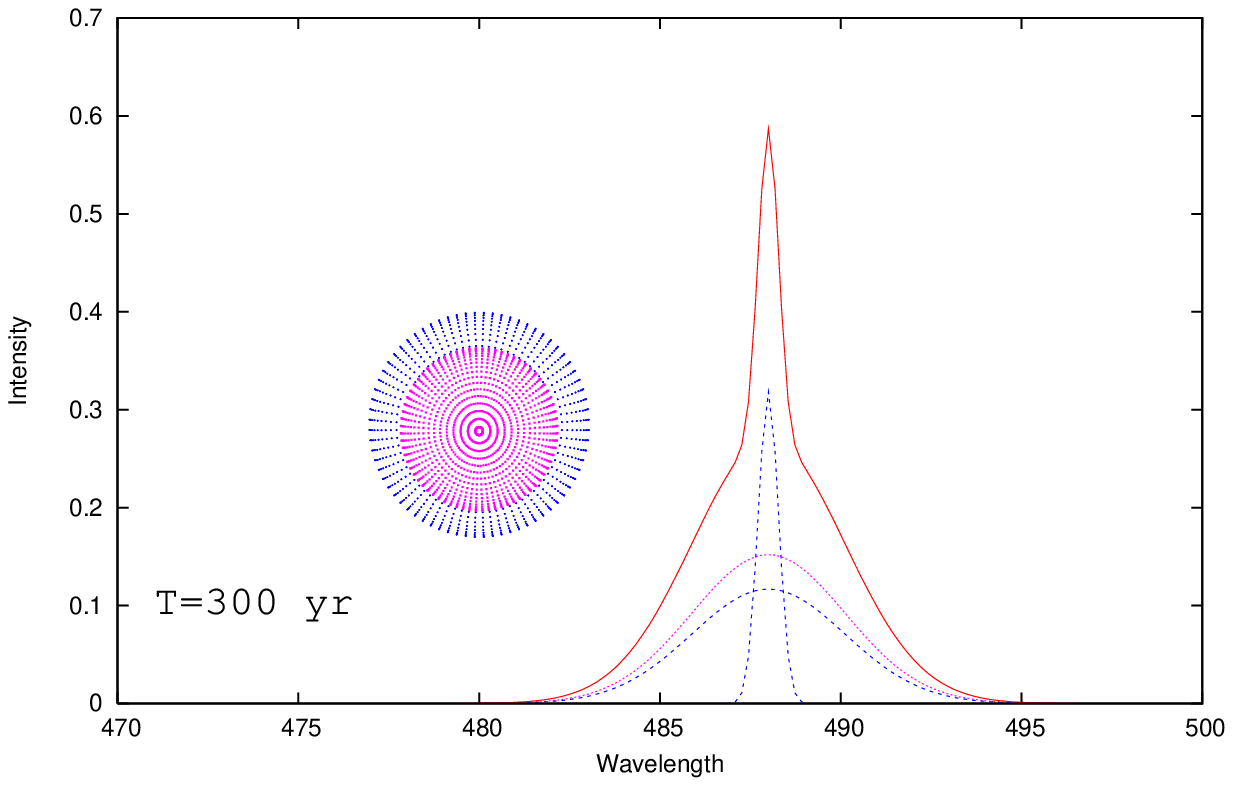}
\includegraphics[width=3.6cm]{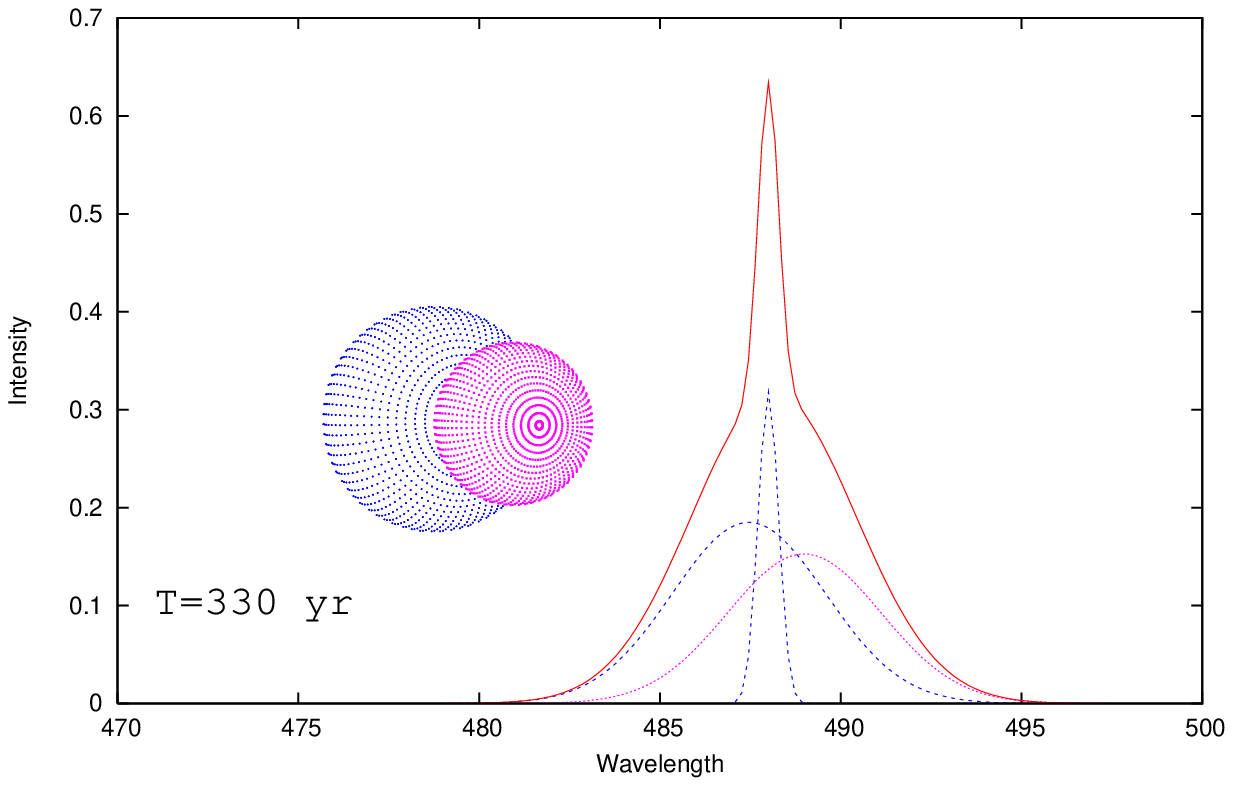}
\includegraphics[width=3.6cm]{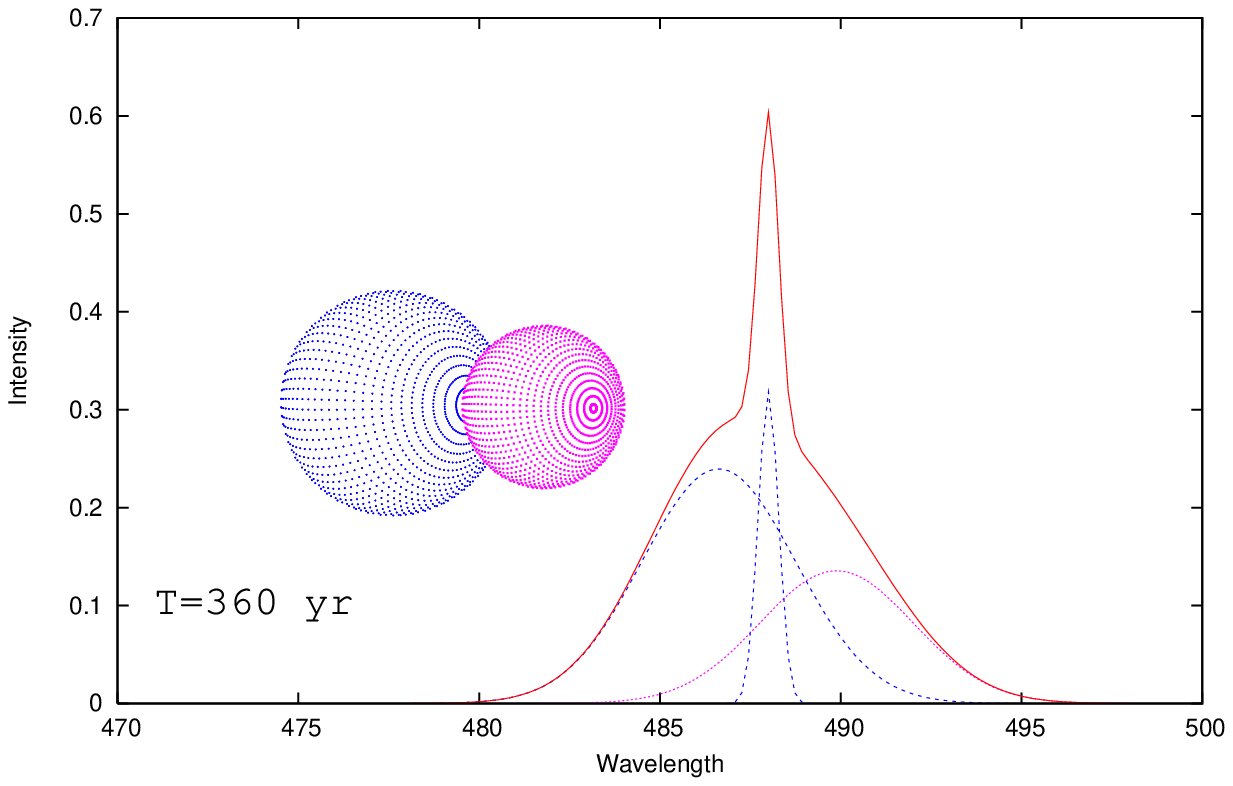}
\includegraphics[width=3.6cm]{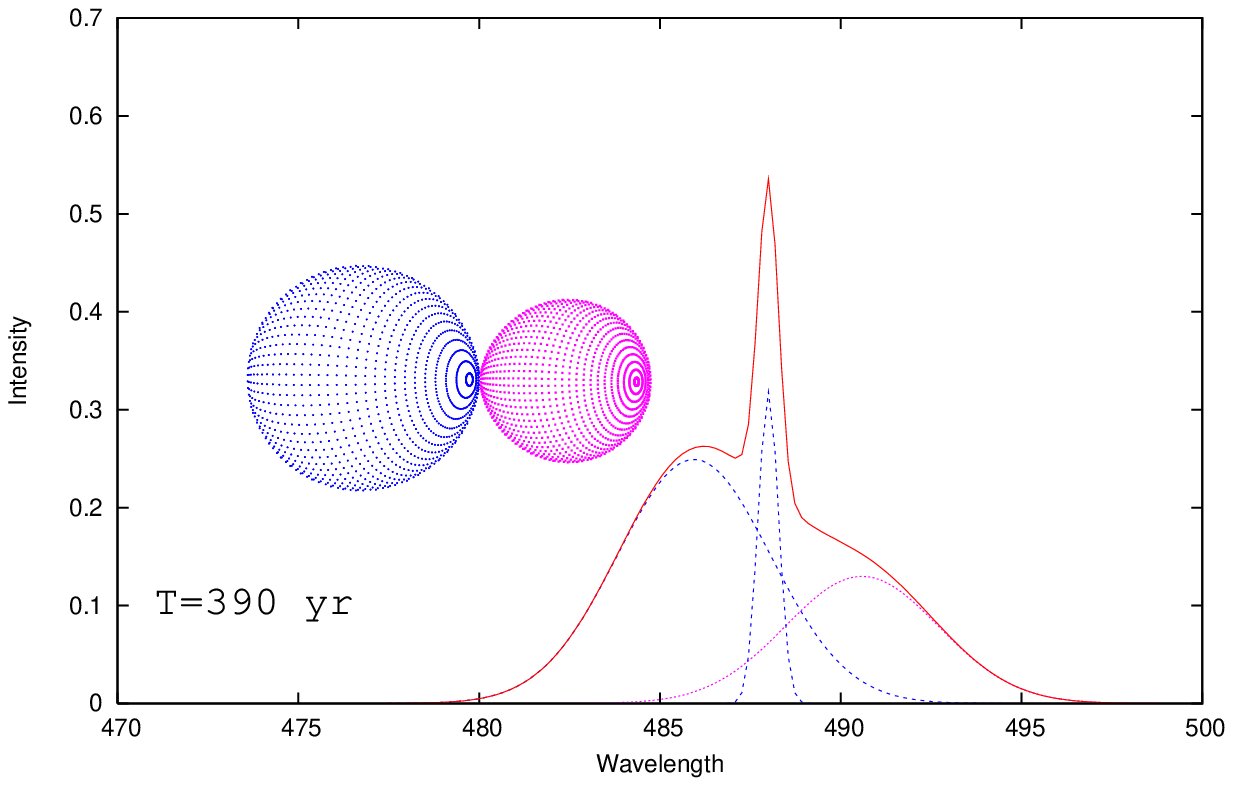}
\includegraphics[width=3.6cm]{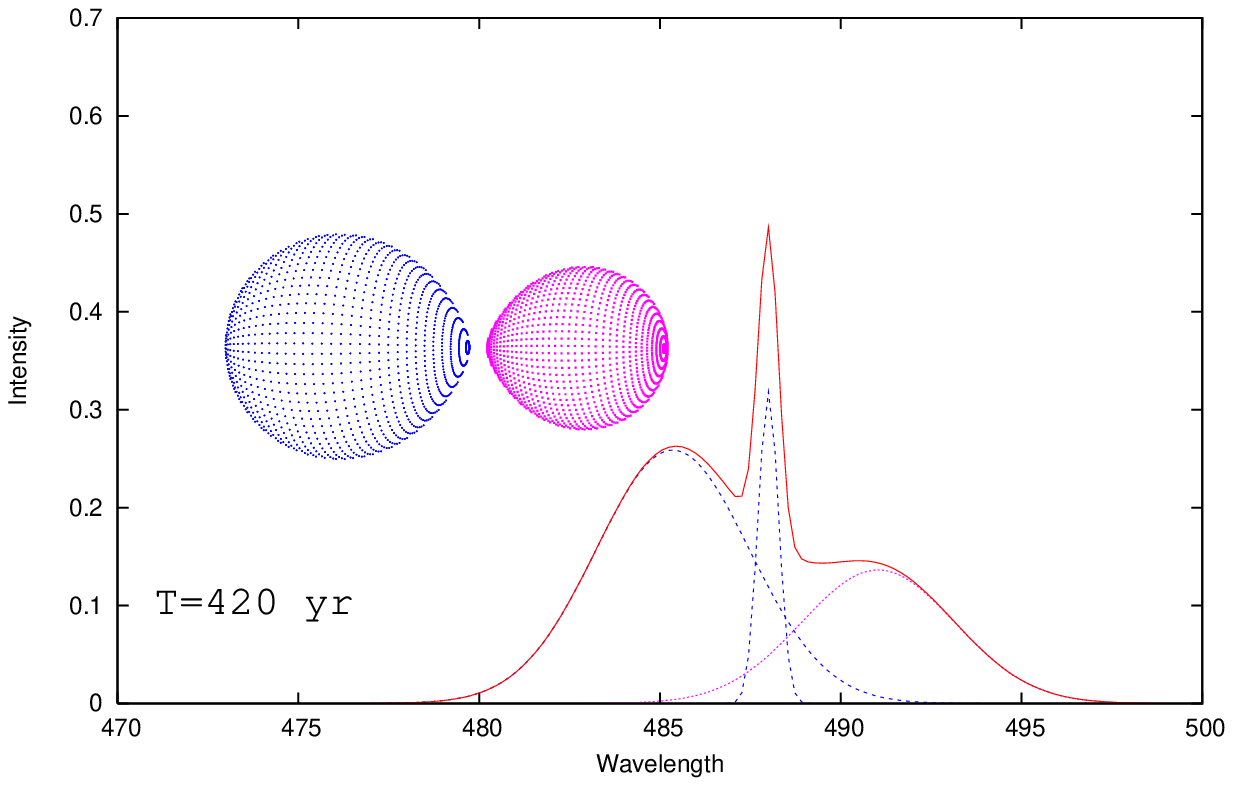}
\caption{The same as in Fig. \ref{f001} first panel, but for different phases of BLRs.}
\label{f002}
\end{figure*}

The emission of binary BLR may result in so called double-peaked broad emission lines in AGNs, therefore such line profiles may indicate the existence of a SMB where two BLRs contribute together to the line profile \citep{g83}, but
as it can be seen in Fig. \ref{f002}, a simple binary BLR model shows that expected broad line profiles can be very different (also
double-peaked). However, a clear double-peaked feature only arises in a particular stage of the binary evolution when the two BHs are close enough such that the line-of-sight orbital velocity difference is larger than the FWHM of the individual broad components (see Table \ref{t01}). Before of this stage, the velocity splitting due to the orbit motion of the binary BLR seems to be very small to be separated and the line profiles are probably asymmetric. In the case that SMB components are closer than dimensions of the BLRs (see Fig. \ref{f-bh}, panels up) the line profile probably becomes more complex and one cannot expect the splitting of the peaks, since it does not correspond to the orbital motion of the binary.
In this case, there are no coherent radial velocity drifts in the peaks with time. Taking all this into account, in the case of a binary BLR one can expect  asymmetric line profiles as a more probable and common signature of binary SMBs than double-peaked profiles.

\subsubsection{Single BLR}

Emission from only one BLR can be observed in two cases: a) There are two SMBHs, but only one has emitting
regions, and b) a recoiling SMBH that has emitting regions.

a) SMBs may evolve within a large-scale gaseous
disk \citep[see e.g.][]{do07,do09a,co09}, and their orbits will be circularized by the time they reach sub-parsec
orbital separations. At smaller separations, gravitational radiation becomes the dominant angular momentum loss process, and any gas trapped inside the orbit of the secondary. SMBs open a hole at the center of the circumbinary disk and the secondary BH orbits closer to the gas reservoir and has easier access to it, i.e., the accretion rate onto the secondary SMBH should be considerably higher than that
onto the primary \citep[see e.g.][]{arm02,ha07,cu09,lo09}. Therefore, it is the best chance that secondary has a BLR. It is in favor of the observation of shift effects on broad emission lines, since (see Eg. (3)), the higher radial velocities can be expected in the case of secondary. In a system where the BLR is associated with the secondary, one can expect shifted observed emission, as is present in Fig. \ref{f001} (panel down).

It seems that
the accreting secondary scenario may be more likely on theoretical grounds.  However, some simulations show that the vast
majority of the gas goes to the big SMBH and therefore one may expect that primary produces the vast majority of  the emitted energy in lines \citep[][]{gas11}.

b) For recoiling BHs the underlying hypothesis here is that the SMBH is fueling by the gas from the (former) gaseous (circumbinary) disk which falls in the SMBH \citep{zan10}. The interaction of kicked SMBHs with the surrounding medium may be, in general, very complex \citep[see][]{go07,zan10}. Accretion rates of the recoiling SMBHs are favorably high for the detection of off-centred AGN if kicked into gas-rich disks - up to a few per cent of the Eddington accretion rate \citep[][]{si11}. Marginally bound gas rejoining the disk around the moving black hole releases a large amount of energy in shocks in a short time, leading to a flare in thermal soft X-rays with a luminosity approaching the Eddington limit. Reprocessing of the X-rays may contribute to strong optical and ultraviolet emission lines with a distinctive spectrum \citep[][]{sbon08}.
Therefore the broad lines are emitted from an, so called, ordinary
BLR that is associated to the recoiling SMBH \citep{go07,zan10}
and that the narrow lines are emitted from the galaxy
where the recoiling SMBH is originated \citep[][]{bonn07}. The narrow lines are illuminated from
the outside (by the recoiling SMBH). The offset AGN lines may be observable either immediately after a high-velocity recoil or during pericentric passages through a gas-rich remnants \citep[][]{b11}. The shift of broad line (with respect to the narrow one) can be expected, and can be of an order of 1000 kms$^{-1}$ \citep[so called kick velocities, see e.g.][]{go07}, or smaller as e.g., \cite{lou10} found that the magnitude of the recoil velocity distribution decays exponentially with mean velocity around 630 km s$^{-1}$ and standard deviation of 534 km s$^{-1}$, that implies probability of 23\%  of recoils larger than 1000 km s$^{-1}$. But the velocities may be significantly higher \citep[see][]{sp11}, i.e there is a good chance to see line shift due to recoiling SMBHs. However, it should be taken into account that one  the projected radial velocity from recoiling SMBH could be smaller, therefore expected line shift is probably around several 100s km s$^{-1}$.
Estimated AGN lifetimes phase of a recoiling SMBH is up to 10 Myr \citep{b11,si11}.

Taking into account discussion above, it seems that observational line profiles emitted from a SMB system probably are asymmetric and shifted regard to narrow lines. The dominant  emission of  single broad emission lines from SMBs (or recoiling SMBHs) seems to be more common than emission of both components. Therefore, shifted broad lines (regarding to the narrow ones) may be an indicator of SMBs presence.

\subsection{Double-peaked and high-shifted broad lines: Complex geometry of the BLR vs SMBs}

The observed AGN broad line profiles are ranging from the classic "logarithmic" profile to double-peaked, disk-like profiles \citep{su00}, and
the problem with using broad emission lines to detect the SMBs is that complex line profiles can be caused by a complex BLR geometry. The emitting gas is relatively close to the SMBH where several processes can affect the line profile. In principle, so called unusually broad line profiles, recently reported in a number of papers \citep[see][]{er11,ts11} are connected with a large shift of broad component ($\sim$ 1000 km s$^{-1}$) regard to the narrow ones or there are strange peaks in the line profiles. Both of these may indicate presence of a SMB system, but also can be explained by complex geometry of the BLR. Here we discuss the so called 'unusual' double-peaked and single(double)-peaked shifted broad emission line profiles in the light of the SMB hypothesis and complex structure of the BLR (and visually close broad line sources).

\subsubsection{Double-peaked emitters}

Different geometries can be assumed in order to describe the BLR \citep[see][]{su00}. As it can be seen in Fig. \ref{f002} binary BLR can result in different line profiles, but similar line profiles can simulated in the case of a BLR complex geometry. There is also possibility that in single AGN, the broad emission lines are composed from two (or more) broad emitting regions \citep[see e.g.][]{pop04,bon09a,bon09b,pop11}, that will result in asymmetric and shifted broad line profiles. The double peaked line profiles seem to be emitted from an accretion disk \citep[see][etc.]{ch89,chf89,er97,st03,pop11}, and also, different effects \citep[as e.g., winds, see][]{mc95} can contribute to the line asymmetry and disk-like profiles. Also, a double-stream model can be applied for line profiles \citep[see, e.g.][]{zsb90}\footnote{Note hera that a simple/pure double-stream model
  has a problem: its only reproduces the line profile but it
  disagrees with the reverberation mapping results for double-peaked
  emitters.}, or combination of the disk emission with double-streams \citep[as e.g., in the case of Ark 120, see][]{pop01}. Also, the off-axis model can explain line profiles from double-peaked (highly asymmetric) to the symmetric \citep{gkn07,gas10}. Here is a problem that physics and geometry of the BLR is not purely understand \citep[see e.g.][]{lcp06}, and consequently, there is a difficulty to separate the influence of SMBs from the complex BLR geometry.
Nevertheless, there are several AGNs that show unusual double-(multi)peaked profiles which have some specific features in the profiles that should be explained. It is interesting to discuss any chance that such, unusual, profiles are emitted from SMBs. Let us discuss several well know 'SMB candidates' which were pointed out because of their unusual double-peaked profiles.

\begin{figure}
\includegraphics[width=8cm]{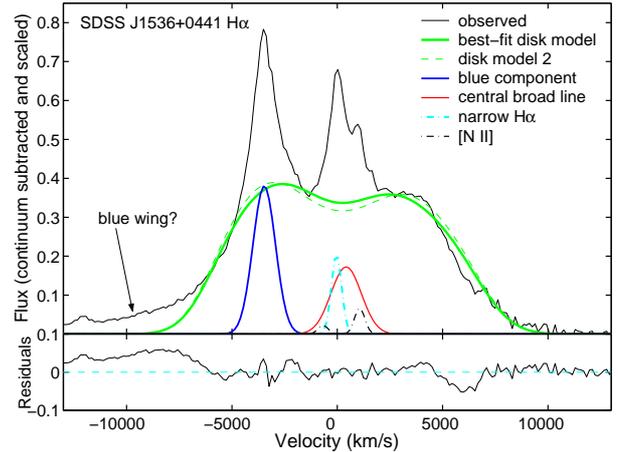}
\caption{The unusual H$\alpha$ emission line profile of SDSS J1536+0441 fitted by the disk model (solid green line)
There are additionally two peaks; left (blue line represented with the Gaussian) and
the central peak (represented with red line Gaussian). Also there is a far blue wing that cannot be fitted with disk \citep{tg09}.
\label{fig-g}}
\end{figure}

There were many discussions about quasar SDSS J153636.22+044127.0 that has peculiar broad emission-line profiles with multiple components and also observed in the X-ray \citep[][]{arz09}. The object is widely discused in several papers \citep{wl09,tg09,bl09,de09,lb09,cho09,wl10,gr10,g10,c10,tw11}. The object shows double-peaked lines, where peak separation is $\sim$ 3500 km s$^{-1}$, and was proposed by \cite{bl09} as a candidate for sub-parsec binary SMB system, but \cite{c10} found in the optical spectra that the system is more likely an unusual member of the class of AGNs known as "double-peaked emitters" than a sub-parsec SMB or quasar pairs. The line profiles were fitted by a disk model by \cite{tg09} and they found that additional emission is present; there is a central blue and central peak in addition to the disk emission (see Fig. \ref{fig-g}). Moreover, \cite{wl09} found from new VLA imaging at 8.5 GHz, two faint sources and they found that the double may be energized here by the candidate 0.1 pc SMB system or the radio emission may arise from a binary system of quasars with a projected separation of 5.1 kpc, and the two quasars may produce the two observed broad-line emission systems. Also, \cite{bp10} presented the pc-scale  radio imaging (with European VLBI Network) of SDSS J1536+0441. at 5 GHz, with the resolution of $\sim$10 mas ($\sim$50 pc). They detected two compact radio cores at the position of the kpc-scale components VLA-A and VLA-B, the compact active nuclei with radio luminosity ${\rm L_R \sim 10^{40} erg\ s^{-1}}$. This observations ruling out the possibility that both radio sources are powered by a 0.1 pc binary black hole. As a conclusion in this case the line emission probably is coming from close projected AGNs (on the kpc scale) and not 0.1 pc scale SMB system. In any case, the unusual broad line profiles of the object stimulate investigations and show that there is a relatively close AGN system.

\begin{figure}
\centering
\includegraphics[height=8cm,width=6cm,angle=270]{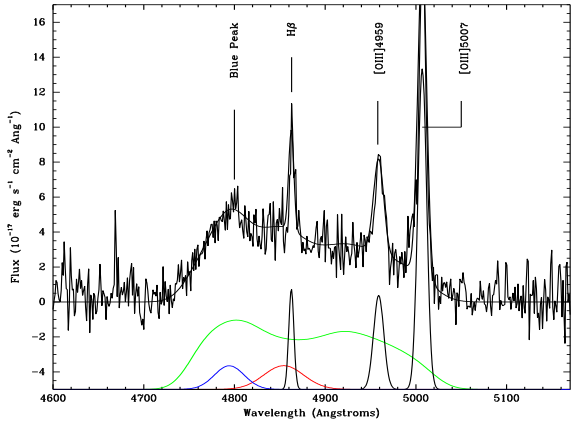}
\caption{Unusual H$\beta$ line profile in J093201.60+031858.7. The line was fitted with disk profile plus best fit Gaussian components \citep[see i][]{ba11}.}
\label{Hb_wdisk}
\end{figure}

However, in some cases the unusual line (double-,multi-peaked) line profiles cannot be exactly explained by the pure (circular) accretion disk emission. As e..g. \cite{ba11} observed very broad (and double-,multi-peaked) line profile in J093201.60+031858.7 (see Fig. \ref{Hb_wdisk}), the strong blue peak is located at 4200 km s$^{-1}$ distance from the narrow line. Note here that SDSS J105041.35+345631.3 has a blueshifted broad component that cannot be explained by the disk emission \citep{sh09}. 
       Additionally, \cite{ts11} found a 'new class' of double-peaked emitters, that are with very broad and faint lines, and one peak is highly shifted ($\sim 5000 \rm\ km\ s^{-1}$).

The common properties of so called unusual double-peaked emitters are that the lines are very broad, and that (at least) one peak is extremely shifted \citep[to the blue, expected in a relativistic accretion disk, see][]{eh94} or there are several peaks in clearly double peaked profiles. As was noted above, such profiles can be explained by the accretion disk emission, but there are some points which has to be taken into account: i) there is extremely shifted (blue) peak (around several thousands km s$^{-1}$ and ii) lines are very broad
\citep[around $\sim$ 10000 km s$^{-1}$, and in some cases around 40000 km s$^{-1}$, see][]{wang05}.
Those properties should be explained by disk emission, and question is: what is special in the disk geometry that produce such lines?

A relatively simple model
\citep[see][]{chf89,ch89} may be used to find some special conditions in  such a disk. First of all, the inclination of the disk should be high
\citep[at least 40-50 degrees, see Fig. 3 in][]{bon09a}, but in this case, it is hard to expect high luminous disk \citep[see][]{bon09a}, and double peaked lines should be faint. Also, very broad line profiles can be emitted from smaller inclinations (30-40 degrees) if the radius of the
  line emitting region of the disk is small \citep{wang05}.  Therefore, the 'new class' of 'extremely double peaked emitters' (weak lines) discussed by \cite{ts11} is probably caused by emission of the high inclined or compact accretion disk. Additionally, there are more sophisticated disk models that are not axisymmetric, as e.g. a disk with a bright spot \citep[the case of 3C390.3, see][]{jov10} or   a disk with spiral arms \citep[see,][]{stb03,st07}. It seems that these
 models can easily fit the observed profiles similar as in  J093201.60+031858.7 without the need to invoke a separate kinematic  component \citep{erac11}

Finally, the broad double-peaked lines seem not to be an appropriate tool for SMBs detection, since for such emission very specific position of the SMBHs
(orbit near edge-on, and both BLR should be clearly resolved) is needed and also a high orbital velocities (which are in principle smaller than FWHM of broad lines, see Table \ref{t01}). Therefore, the disk, or disk-like geometry probably cause the double-peaked profile, and 'unusual' weak double-peaked broad lines \citep[][]{ts11} probably come from the high inclined disk.

\subsubsection{High off-center shifted broad emission lines}

The high off-center shifted broad emission lines may indicate two stage of a SMB: i) the case where in SMBs only one  component has a BLR, and ii) emission of the BLR from a recoiling SMBHs, i.e., stage after coalescence. There are several AGNs with a relatively large shift of the broad line component \citep[see e.g.][]{er11,ts11}, between them SDSS J092712.65+294344.0 that was the first candidate for a recoiling SMBH \cite{kz08}, with the blueshifted broad component of 2650 km s$^{-1}$ relative to the narrow emission lines.

 First \cite{g83} discussed the displaced emission lines in light of SMB hypothesis and reported about two quasars \citep[see Fig. 1 and 2 in the paper of][]{g83} that broad lines in 0945+076 and 1404+285 were off-centered (shifted) to -2100 km s$^{-1}$ and +2700 km s$^{-1}$, respectively

\begin{figure}
\centerline{
\includegraphics[]{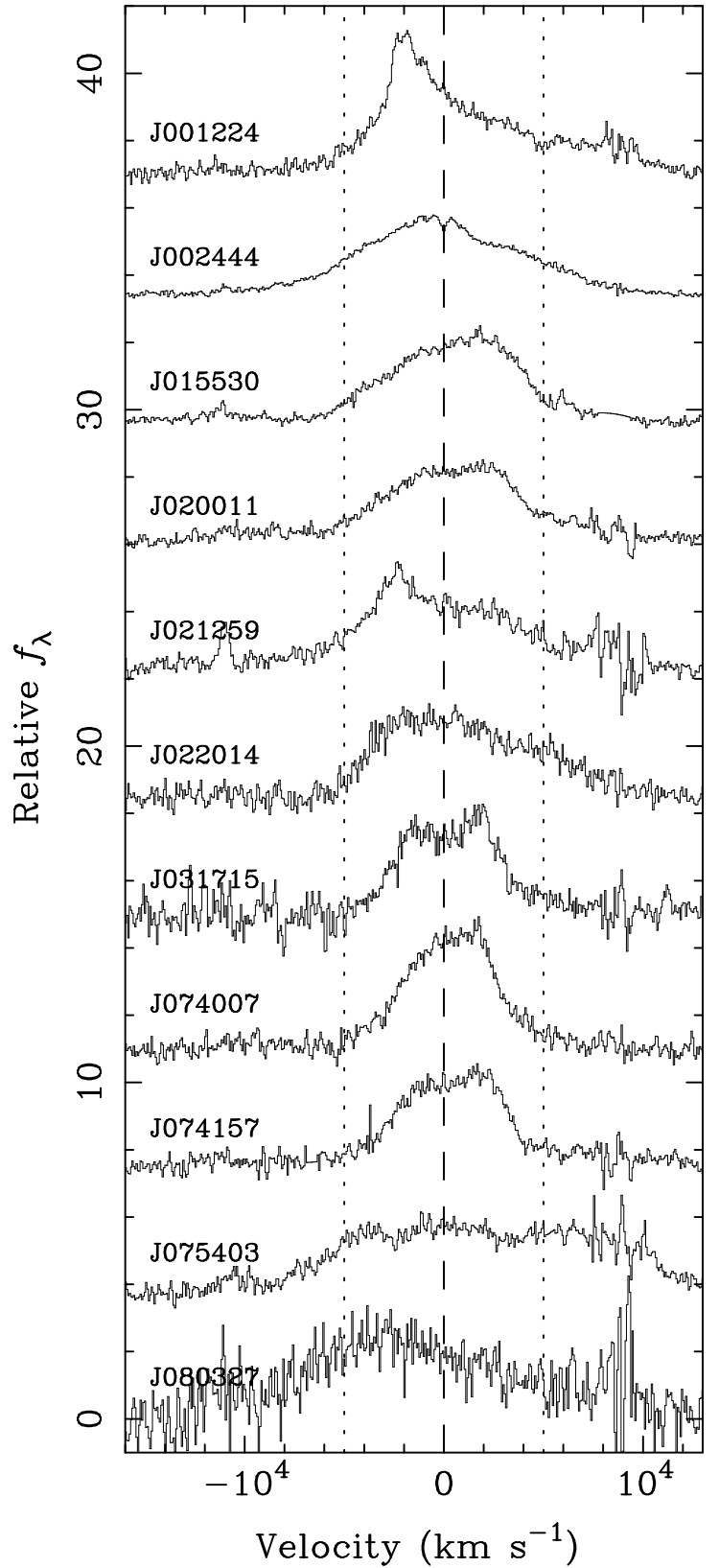}
}
\caption{Unusual line profiles of the broad H$\beta$ from the
 SDSS spectra reported by \citep[][]{er11}. The solid vertical line denotes the H$\beta$ narrow line location.
The vertical dotted lines indicated a
 window of $\pm 5,000$ km s$^{-1}$ the center \citep[see][in more details]{er11}.
\label{f-er11}}
\end{figure}

Several other candidates are worthy to mention 4C+22.25 \citep{de10a} where the H$\beta$ and H$\alpha$ lines show very broad line profiles (FWHM $\sim$ 12,000 km s$^{-1}$), faint fluxes, and extreme offsets (8700$\pm$ 1300 km s$^{-1}$) with respect to the narrow emission lines, but the line profiles do not vary in a period of 3.1 yer \citep{de10a}. In E1821+643 the broad Balmer lines are redshifted by $\sim$1000 km s$^{-1}$ relative to the narrow lines and they have highly red asymmetric profile \citep[][]{rob10}. In polarized flux the broad H$\alpha$ line exhibits a blueshift of similar magnitude and a strong blue asymmetry. \cite{rob10} showed that these observations are consistent with a scattering model in which the broad-line region has two components, moving with different bulk velocities away from the observer and toward a scattering region at rest in the host galaxy, and they concluded
 that the SMBH  itself is moving with a velocity $\sim$2100 km s$^{-1}$ relative to the host galaxy that very well fit the recoiling SMBH hypothesis.
In the case of SDSS J105041.35+345631.3 a large velocity blueshift (3500 km $\rm s^{-1}$) and a symmetrical H$\beta$ profile make this object an interesting candidate for black hole recoil \citep{sh09}. Additionally, there is a group of such objects recently discovered by \cite{er11} and \cite{ts11}. The shifted broad line profiles of several SDSS AGNs discovered by \cite{er11} are shown in Fig. \ref{f-er11}.

Let us here consider in more details the recoiling black hole candidate SDSS J0927+2943 \citep[first noted candidate][]{kz08}, that has been observed and discussed in several papers. The quasar was suspected to be kicked off \citep{kz08}, and second idea was that it is hosting a massive black hole binary embedded in a circumbinary disk. \cite{do09b} found that the system is the SMB with mass ratio q $\sim$ 0.3, where the mass of primary is $\sim 2 \cdot 10^9 M\odot$. The semimajor axis was estimated as 0.34 pc, corresponding to an orbital period of 370 years. They found that the binary hypothesis is preferred being 100 times more probable than the ejection hypothesis.

However, the velocity difference between two AGNs is $\sim$ 2500 km s$^{-1}$ that is, according to \cite{dr10}, in favor of superposition of two AGNs at different redshifts. Namely, \cite{dr10} re-examined the superposition model and showed that AGN pairs with high velocity line separations up to $\sim$2000 km s$^{-1}$ are very likely to be superpositions of two AGNs, but no superimposed AGN pairs are predicted for velocity offsets in excess of $\sim$3000 km s$^{-1}$.

Moreover, the long-slit spectroscopic observations of SDSSJ092712+294344 showed two systems (quasars), one, at z = 0.7128, being at an impact parameter of $\sim$1 kpc with respect to the another quasar at z = 0.6972 in the north-west direction \citep[][]{viv09} and it seems that the SMB model is most unlikely. Additionally, \cite{der10b} discovered a quasar close to SDSS J0927+2943, that has a radial velocity difference of $\sim$1400 km s$^{-1}$ and projected distance of 125 h$_{70}^{-1}$ kpc\footnote{125 kpc for H$_0$70 km s$^{-1}$Mpc$^{-1}$}. Also \cite{sh09a} supported the idea that this system is a superposition of two active galactic nuclei. Therefore, off-centered broad line profiles in this object may be explained as an emission of two visually close binary quasars, with small angle separation.

In general, line shape of SDSSJ092712+294344 (and other similar emitters) may indicate four possible situations: i) a recoiling SMBH \citep{kz08}; ii) a SMB where one BLR emits \citep{do09b}; ii) a superposition of two visually close, (non-)interacting, AGNs \citep{viv09,der10b} and iv) an effect of the complex velocity field in the BLR.

\begin{figure}
\includegraphics[width=6cm,angle=-90]{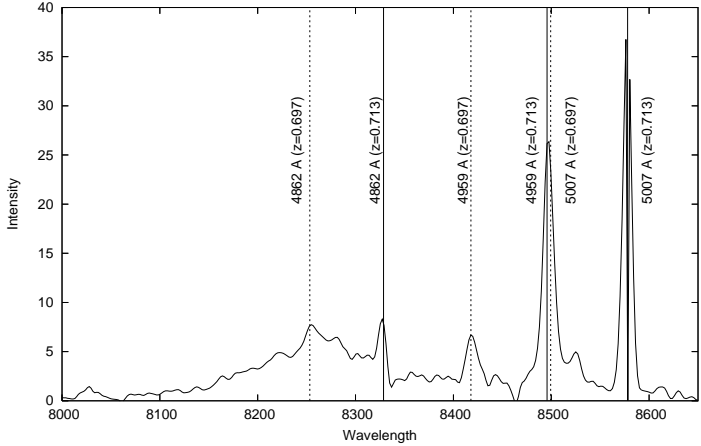}
\caption{The observed H$\beta$ profile of SDSS J092712.65+294344.0. The vertical dashed lines correspond (from left to right) to the H$\beta$ and [O\,III]$\lambda\lambda$4959,5007 with the cosmological redshift z=0.697 and solid vertical lines to H$\beta$ and [O\,III]$\lambda\lambda$4959,5007 with z=0.713. This and difference in FWHM between [O\,III]$\lambda\lambda$4959,5007 (with z=0.713) lines and their smaller flux ratios ($<2$, expected $\sim$3) ruled out recoiling SMBH hypothesis and confirmed visually close AGNs.
\label{fig-sdss}}
\end{figure}

 Additionally we analyzed in more details the spectrum of SDSS J092712.65+294344.0. Using a simple analysis we
try to identity the lines, and found that there are probably merged emission of two AGNs (see Fig. \ref{fig-sdss}) one with redshift of z=0.697 (denoted with vertical dashed lines in Fig. \ref{fig-sdss}) and one with z=0.713 (vertical solid lines). From a simple analysis we found two redshifted sets of lines almost the same as in \cite{viv09}. It is interesting that [O\,III]$\lambda$4959 of one object, at z=0.713, is superposed with [O\,III]$\lambda$5007 of another object at z=0.697. Additional supports of the hypothesis of emission of visually close AGNs at different redshift is the ratio of the [O\,III] lines that is smaller than 2 \citep[but should be around 3, see][]{dim07} and that there is significant difference in the [O\,III] widths, shifted at z=0.713 where [O\,III]$\lambda$4959 \AA\ is significantly broader (400 km s$^{-1}$) than [O\,III]$\lambda$5007 \AA\ ($\sim$270 km s$^{-1}$).

 As we mentioned above, broad high off-centered line profiles can be explained by different effects in the BLR. Such profiles may be explained by the
distribution of BLR gas proposed by \cite{gkn07,gas10}, as e.g., in Fig. \ref{f00} (up) is an example of the unusual, H$\beta$, profile of SDSS J0946+0139 (dots) and the best fit with model given in \cite{gkn07}. As one can see from the Figure, the model can perfectly explain 'unusual' profile of this object \citep[see in more details][]{gas10}. There are other models, as e.g., two component model, taking high velocity (out-)in-flow \citep{pop04,bon09a,bon09b} can reproduce similar line profiles. Additionally, as it was presented in \cite{il10}, an accelerating outflow model for the BLR (very close to the SMBH) can generate broad emission line profiles ranging from the double peaked to the high off-center shifted broad lines.

Nevertheless, it is not possible to totally rule out the SMB scenario in the case of unusual and off-centered broad line profiles, as e.g., SDSS J1529+333 \citep{ts11} where the peak shift is extremely high ($\sim$6000 km s$^{-1}$) and also stellar distribution in the host is extended, indicating perturbation in the stellar disk structure \citep[][]{ts11}.
Generally, the recoiling SMBH hypothesis had small probability in these cases because of high off-center shift \citep[although kick-off velocity may be extremely high in some special cases, see][]{sp11}. Next possibility is that here may be a SMB system with significantly smaller mass of a secondary that has a BLR. Future observations and additional evidence will confirm or rule out the SMB scenario for this object.

At the end considering a probability that SMBs emit single high (low) shifted or double-peaked lines, it seems that is higher probability to observe single low (high) shifted lines than double-peaked in such a system, however, the problem to resolve the SMB effect and effect of the complex BLR structure remains. As one example, we show in Fig. \ref{f00} (down) similar (off-center shifted) line profile that may be emitted from a SMB system with a binary BLR, and observations can be fitted with specific model of the BLR \citep[see panel up, and][in more details]{gas10}.

\subsection{Variability of the broad line profiles and SMBs}

The variability of broad spectral line shapes is in principle expected in the case of a SMB system \citep[see][]{g83,g85,g96a,g96b}. First, the variability
can be caused by orbital motion of the SMBs, i.e., different positions of the component with regard to the observer.

\begin{figure}
\includegraphics[width=7.5cm]{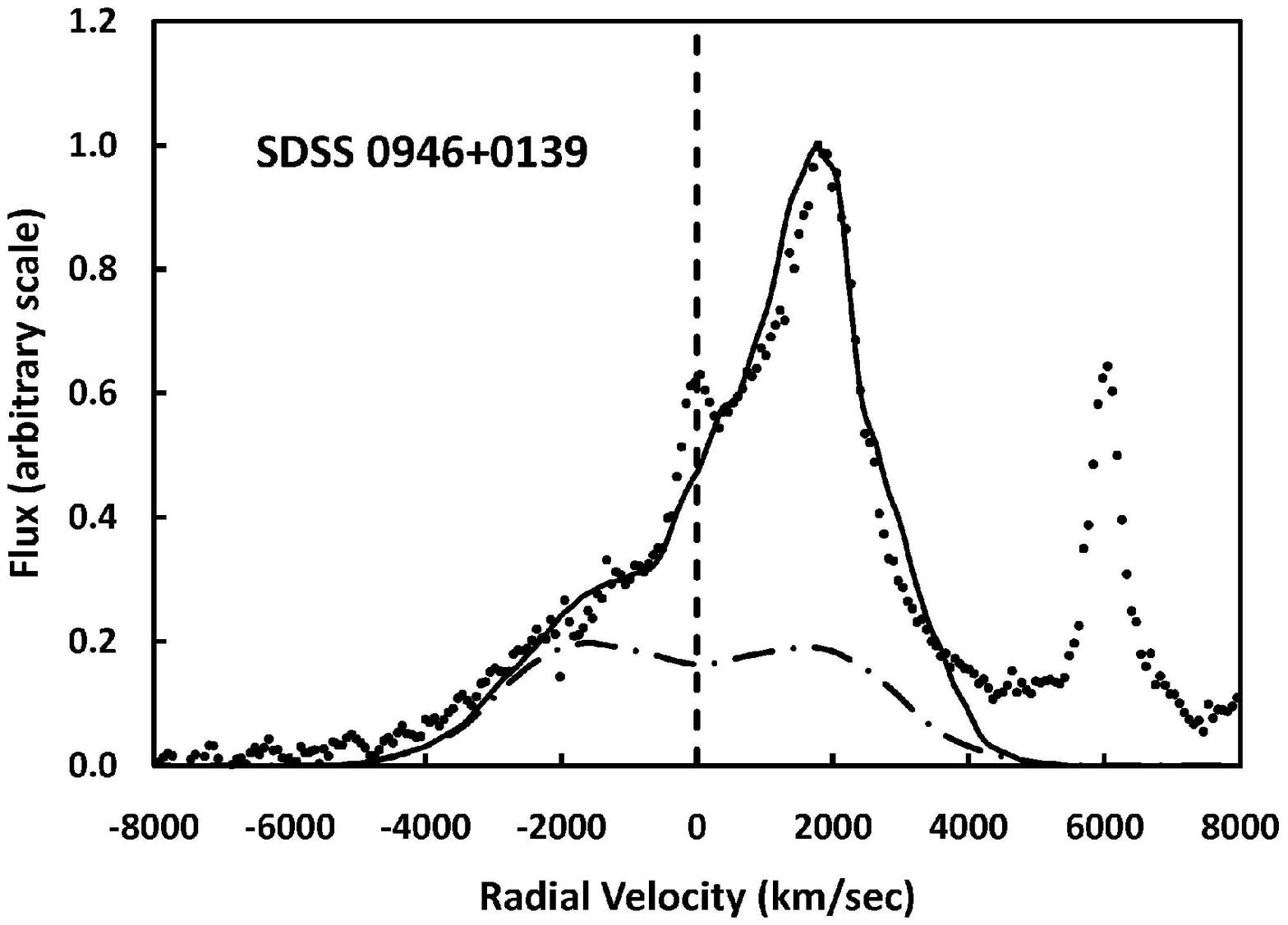}
\includegraphics[width=8cm]{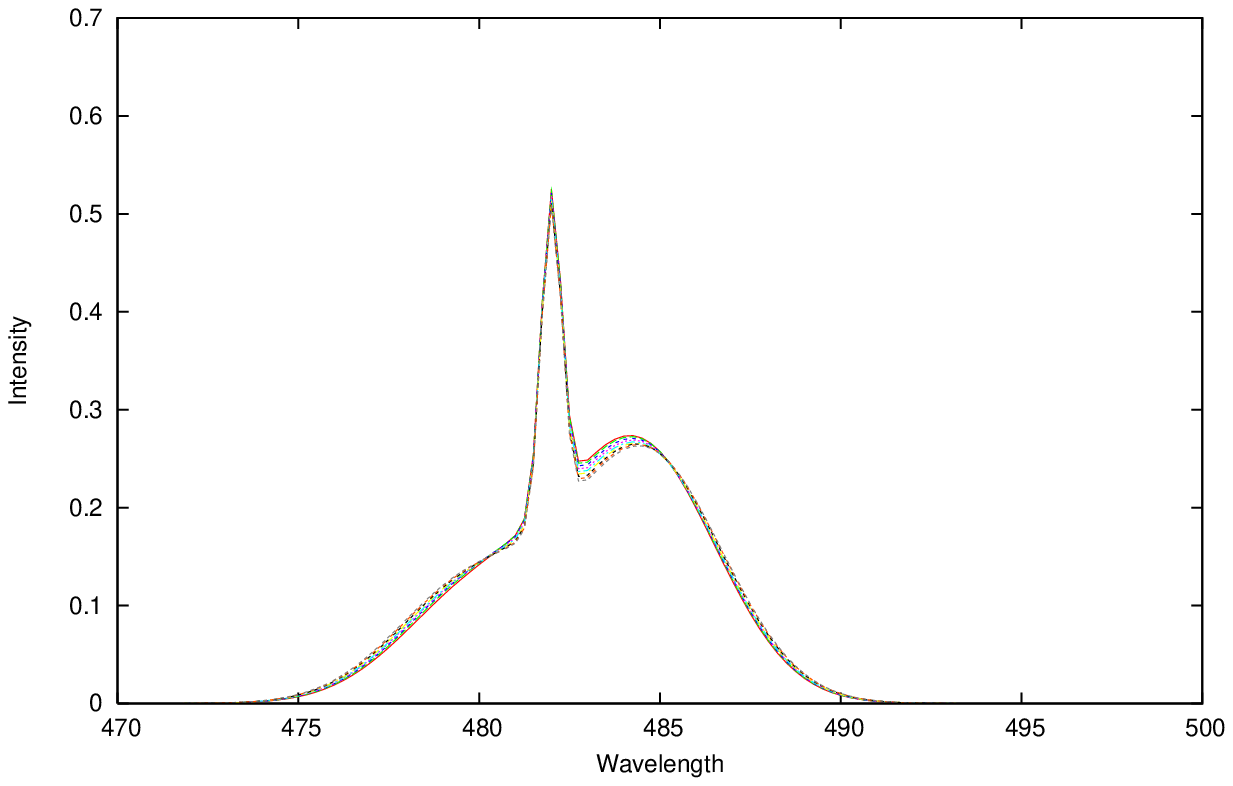}
\caption{Up: The observed, unusual, H$\beta$ profile of SDSS J0946+0139 (dots) compared with the off-axis illumination
model \citep[solid line, model is given in][]{gkn07,gas10}. Down: Similar profile, obtain by a toy model \citep[for the SMB system described in\S 3][]{p00}. The variability due
to different phase (position with regard to the observer) in the 11-year period are simulated (seen as slight changes in the left and right side from the
narrow line.
\label{f00}}
\end{figure}

 We explored variability of the line profile due to orbital motion of the binary using above mentioned toy model \citep[][]{p00}. We obtained that in a
11-year period,
see Fig. \ref{f00} (down), there cannot be detected significant variability in the line profile due to the orbital motion of binary at $>0.1$ pc distance (see Eq. 2). That is expected in the case of close BLR, since if there is a BLR around each of the SMBHs, and system is not semi-detached, the orbiting periods are probably around a couple of 100s years.

 On the other hand, the broad lines show variability in many AGNs, and one can expect that the broad line regions in a SMB (both of them) may show variability, and in
this case the broad line profiles should change during the time. The expected variability is that the far wings will not change simultaneously. It may be the case where projected orbital velocity is high, and the broad emission lines are  separated (or double-peaked). However, as we noted above (see also Table \ref{t01} and Fig. \ref{f002}), the most probably is that line shapes emitted from a SMB are asymmetric, and drift velocity is probably smaller than FWHM of each component. This will lead that both (blue and red) wings vary nearly simultaneously.

Considering different phase in SMB evolution, the variability can be significant and important when SMBHs are closer than characteristic distance for the BLR (see Table \ref{t01}), i.e., in the case of semi-detached SMB system. In this case one can expect merging of the gas from two BLRs, and different shocks in a new formed BLR. This may produce a high variability, not only in the intensity of the broad lines, but also in the line profile, since orbiting period is getting shorter, and geometry of the emtting gas is driven by the orbital motion of the SMBHs and thermodynamical processes.

 Additonally,  nonaxisymmetric waves can be induced in the inner part of the circumbinary
disk by the tidal potential of the binary \citep{ho09}. These waves could cause changes in emission line profiles.
Therefore, variability in the line profiles and intensity may be very perspective for discovery of sub-parsec SMBs.

However, a line variability is very frequent in the broad lines of AGNs, and the question is: what is the expected difference between the line variability of the BLR around one SMBH and around a SMB?
First of all, in the case of a single SMBH, the variability is caused by the brightness of the central source, and one cannot expect huge and violent variability in the line shape. Contrary, in the case of semi-detached BLR, geometry may be affected by waves and shocks caused by tidal potential of the SMBHs. Therefore, AGNs showing unusual and violent changes in the broad line profiles  potentially may be candidates for SMBs. Note here that quasars very often show variability in the broad line spectra, an amplification in the line intensity as the response to outburst in the continuum source emission \citep{gas86,pg86,pe04}. But some AGNs, additionally to the line flux variation, show a very high variability in the line shape, as e.g., Mrk 926 \citep[][]{koz10} where blue wing shows higher variation than red one. Another example is, well known AGN, NGC 4151 that shows very huge changes in the line profile during an 11-year monitoring period \citep[see][]{shap10}. However, a problem is that such violent changes in the line profile can be caused by other effects, as outflows, inflows, perturbation in the emission disk, etc. \citep{gas10,shap10,jov10}.

\subsection{Probability of SMBs detection using broad emission lines}

To find the probability to observe a binary BLR, one should clarify how often SMBs are present in the Universe. There are some estimates, as e.g., \cite{hk02} predicted that less than 10\% of faint ellipticals and 40\% of bright ellipticals harbouring an SMB in their center with near equal masses, while \cite{v09}, usinng models of assembly and growth of SMBHs in hierarchical cosmologies, found that sub-parsec quasar binaries are intrinsically rare, i.e., the best models predict ~0.01 deg$^{-2}$ sub-parsec binary quasars with separations below 10$^4$ Schwarzschild radii (with corresponding orbital velocity larger than 2000 km s$^{-1}$ at z $<$ 0.7). In a sample of 10,000 quasars, they predicted an upper limit of only 10 sub-parsec binary ones that gives 0.1\%, but the number of binaries increases rapidly with increasing redshift and one can expect more SMBs at larger redshift.

As we mentioned above, the mass ratio is very important in the dynamics of the SMBs, and the investigation performed by
\cite{ha10} showed that (1.5\%$\pm$0.6\%) of the total number of nearby AGNs for the equal-mass ratio and (1.3\%$\pm$0.5\%) for the one-to-ten mass ratio have close binary massive black holes with an orbital period of less than 10 yr. Note here that such objects may be detectable with ongoing highly sensitive X-ray (see Table \ref{t01} and discussion further in the text).

It is interesting if all SMBs have the equal-mass ratio then one can expect that about 10\% of AGNs with black-hole masses of 10$^{6.5-7} M\odot$ have close binaries \citep{ha10}. This provides a good chance to detect such systems.

Concerning recoiling SMBHs, there is a small probability to observe them. \cite{bonn07} found an upper limits on the incidence of recoiling black holes in QSOs of only 0.2\% for kicks with velocities greater than 800 km s$^{-1}$, and significantly smaller for higher velocities (e.g., 0.08\% for kicks greater than 2000 km s$^{-1}$).

The probability to observe BLR from SMBs derived from observations of broad emission lines are significantly lower. E.g., \cite{ts11} presented the results of a systematic search for massive black hole binaries in the SDSS spectroscopic database, taking some constraints \citep[see][]{ts11} and for a sample of 54586 quasars and 3929 galaxies at redshifts $0.1<z<1.5$ only nine possible candidates for SMBs are selected\footnote{Note here that they found 32 objects with peculiar broad spectral lines }. Some better score of 14 candidates was find in \cite{er11}. This give an experimental result of an order of 0.02\% candidates from the SDSS datebase. Such difference between predictions and discovered SMB candidates may be explained with the fact that the broad emission line profiles (emitted from binary SMBHs) are mostly asymmetric and with a small radial velocity drift (see Fig. \ref{f002}). This may indicate that we observe a number of binary SMBHs supposing that they are ordinary AGNs. This could have some consequences in the BLR physics investigation, as well as in estimates of SMBH masses using broad emission lines \citep[in more details, see review][in this issue]{ms11}.

Note here that in the radio emission there is a slightly better chance to see radio double-core sources. \cite{bu11} searched for binary supermassive black holes using a radio spectral index mapping technique which targets spatially resolved, double radio-emitting nuclei. Only one source was detected as a double nucleus in a sample of 3114 radio-luminous active galactic nuclei, that is around $\sim$0.03\%.

\section{Double peaked narrow lines and SMBs}

Here we will shortly discuss the narrow lines and SMBs. Taking into account that the
 narrow line region -- NLR is the most extensive region in AGNs, the galaxy mergers (or binary black holes) at a larger distance may be detected by the narrow
emission lines (NELs). It is better to call such systems the interacting AGNs. Interacting AGNs at distances of several kpc may show two sets of narrow emission lines in a single spectrum owing to the orbital motion of the binary, but there is a problem with the radial velocities, since (see Table \ref{t01}) an expected radial velocity is significantly smaller than observed FWHM of the narrow lines.

Nevertheless, using SDSS database \citep[see e.g.][]{s07}, it was found that in a number of AGNs the double-peaked narrow lines are present \citep[see e.g.][]{x09,w09,sm10,li10a,li10b,c10,peng11,fu11,f11}.

The double peaks of NELs could result from pairing active SMBHs in a galaxy merger, but also could be due to bulk motions of narrow-line region gas around a single SMBH \citep[see e.g.][]{r10}, and also possibility of projected close visually AGNs at different redshifts.

There are several examples where double-peaked narrow lines indicate interacting AGNs: \cite{f11} observed a SDSS sample AGNs with double peaked NELS (total 50 AGNs, 17 type-1 AGNs between $0.18 < z < 0.56$ and 33 type-2 AGNs between $0.03 < z < 0.24$) and found that the new images reveal eight type-1 and eight type-2 sources which are apparently undergoing mergers. Another example is SDSS J09527.62+255257.2, a radio-quiet quasar shown by previous imaging to consist of a galaxy and its close (1.0$^{''}$) companion. \cite{mcg11} found two AGNs, one of a Type 1 AGN with both broad and narrow lines, and a type 2 companion galaxy with only narrow lines. These two AGNs are separated by 4.8 kpc, i.e., they are interacting galaxies. Moreover, \cite{w09}
found a good correlation between the ratios of [O III] shifts and the double peak fluxes in a sample of 87 AGNs from the SDSS, that  favors the explanation of kpc-scale binary AGNs.

However, it seems that the most of double peaked NELs are originated in the complex NLR than in kpc-scale SMBs
\citep[see][]{s11}. Additionally, \cite{tw11} observed in the radio band of a subsample of 11 AGNs taken from the sample of 87 double peaked narrow line objects. They did not find any double cores, this does not substantially support the idea that AGNs with double-peaked optical emission lines contain binary black holes. There are processes which can result in double peaked narrow lines \citep[as e.g., outflow, or systematic motion of gas, etc., see][]{s11}, but they can be product of the inner structure NLR \citep[][]{s11} and emission in the same lines that comes from host galaxy \citep[see e.g.][]{cr10}. It seems that in principle, a small number of double-peaked NELs have kpc-scale SMB origin, as e.g., \cite{fu11} estimated that only ~1\% of binary AGNs would appear as double-peaked NELs in SDSS spectra.

At the end let us discuss the possibility to detect a recoiling SMBH and sub-parsec SMB using NELs.

In the case of a recoiling SMBH, the accretion disk, formed around kicked off SMBH, is illuminating the gas \citep{si11,fa11} in its former host
galaxy from the outside. The narrow emission lines arise from gas predominantly orbiting in the potential of the host galaxy and will not follow the recoiling SMBH \citep[][]{me06}. The recoiling SMBH may also ionize the gas from host galaxy, producing narrow emission lines, and forming an NLR different from one emitted by the NLR of host galaxy. In this case one should expect complex line ratios and widths, and split between in the narrow lines.

\begin{figure}
\includegraphics[width=4cm]{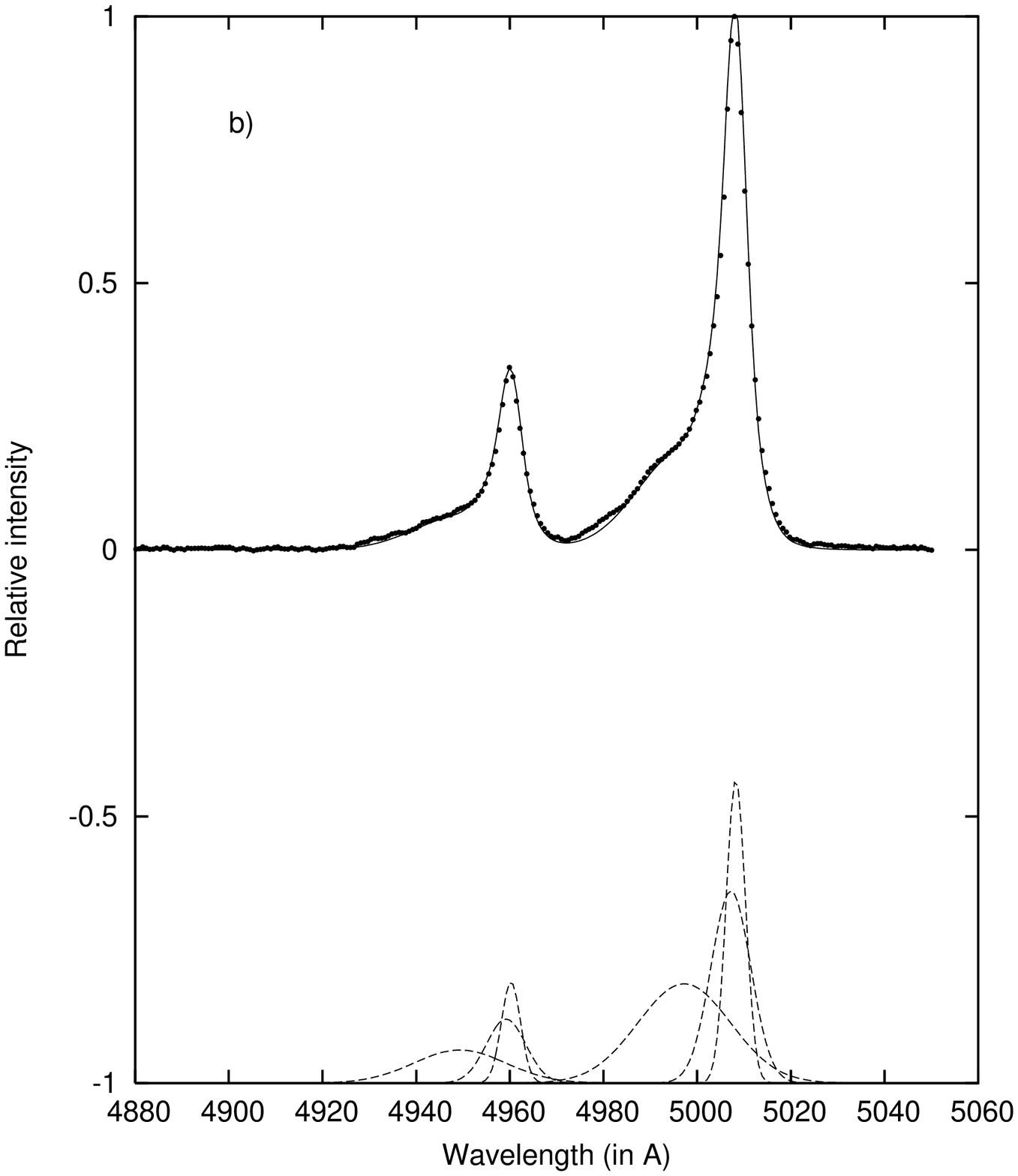}
\includegraphics[width=4.5cm]{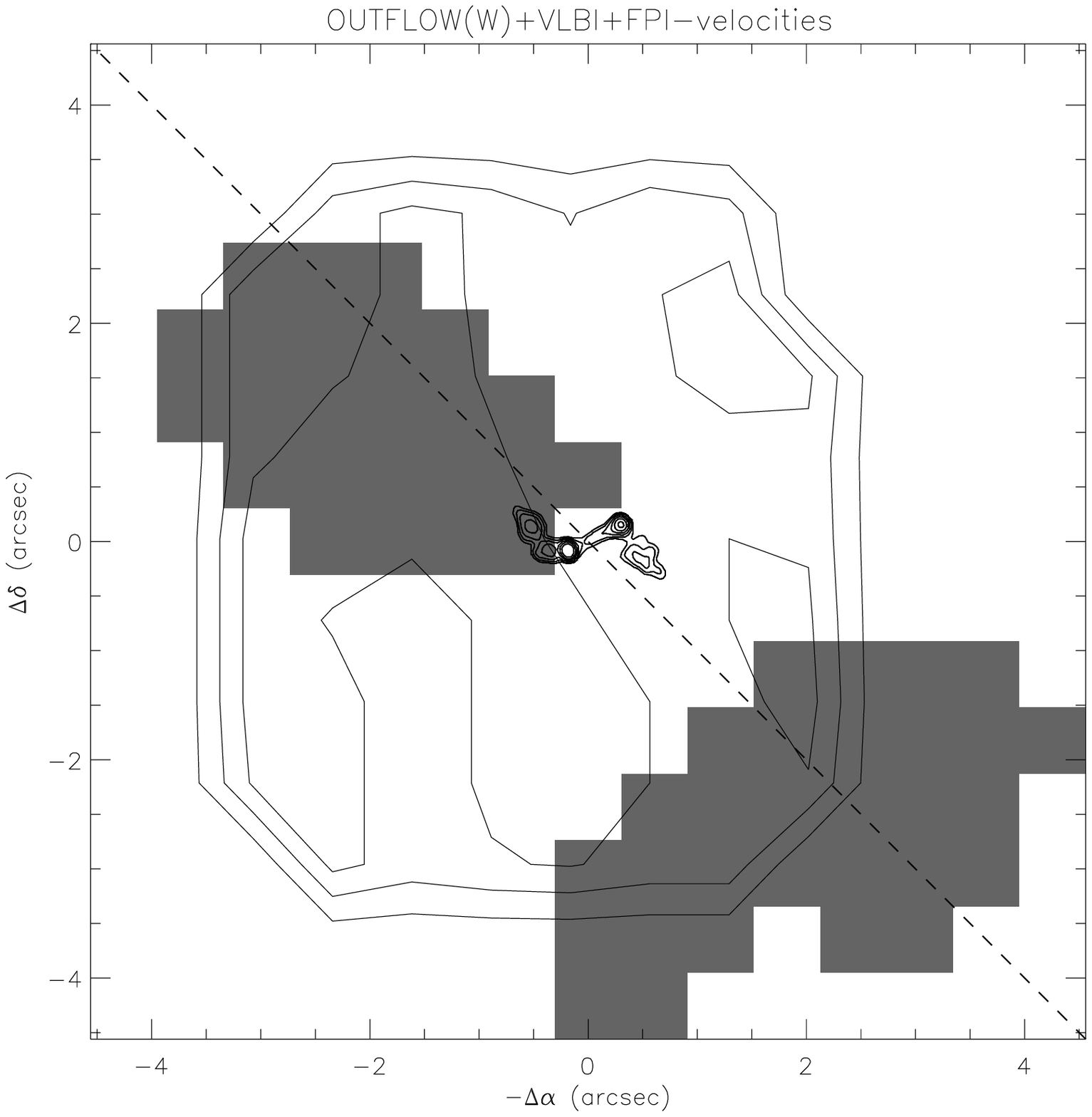}
\caption{The complex narrow line profiles of Mrk 533 (up), it seems that there are three different velocity regions and superposition of the velocity fields obtained from the [O\,III] lines. Down: The outflow map \citep[gray columns][]{sm07} and radio observations of the S-shaped radio structure \citep{mom03} that may indicate a SMB system in the center of Mrk 533.
\label{f-sm}}
\end{figure}

On the other hand, if a SMB system is embedded by the low density gas, the gas (NLR) will be illuminated by two sources, and a high anisotropy in the NLR is
expected. This can lead to different line ratios than in the case where gas is illuminated by only one source. Additionally, perturbation in the gas inside the NLR (where orbiting of the binary is expected) may affect the kinematics of the outher part of the NLR, i.e., a complex velocity field in the NLR may be present, and consequently it would result in the complex narrow line shapes. Note here the case of Mrk 553, where VLTI radio observations of the central part recognized overall S-shaped pattern that (beside of compact radio jet in the center) might be a black hole merger \citep{mom03}. The investigations of the [O\,III] kinematics showed a very complex NLR structure, and it was interesting that velocity field very well corresponds to the S-shaped structure \citep[see Fig. \ref{f-sm}, and paper of][]{sm07}.

The double peaked narrow lines, probably, in a few cases may indicate kpc-scale SMBs, but more perspective seems to be to investigate the influence of the pc or sub-pc SMBs on the narrow line gas physics and geometry. An anisotropy in the NLR is expected in that case, and might be that the [O\,III]$\lambda\lambda$ 4969,5007 lines (their ratio and shape) could give more indication that the source of illumination of the NLR is not a single SMBH.

\section{The Fe K$\alpha$ line and SMBs}

The broad iron line (Fe K$\alpha$), at 6.4 keV at rest energy, is observed in a number of AGNs \citep[see][]{n97,n07,l10}, and it is assumed that the line is emitted from a compact relativistic disk around a SMBH. The emission can come from the first stable orbit and consequently can be used to explore the basic SMBH properties \citep[see in this volume][and references therein]{j11}. There are a number of papers discussing the possibility to use the Fe K$\alpha$ line for the diagnostic of black hole parameters \citep[see, e.g.][etc.]{jov08,pj09,bal10,pat11,j11}.

As it was noted above during of merger phase of two SMBHs there is a sufficient amount of gas  with some angular momentum that it is accreting onto the SMBHs in the form of accretion disks that emit strong X-ray radiation. It was found that the high-energy radiations are emitted from an accretion disk around each black hole long before the black hole coalescence \citep[][]{one09,ha11}. In the case of recoiling
SMBH, it is expected that gas falls in the SMBH producing high X-ray luminosity that could exceed $10^{45}\rm\ erg s^{-1}$ \citep[see][]{fuj09}.
In the case of X-ray SMB emitting disk(s), the two accretion disks may be warped at outer parts and connected with an outer large circumbinary accretion disk \citep[see][]{yl01}. This can indicate that (one or) two accretion disks can contribute to the X-ray radiation, and consequently to the Fe K$\alpha$ line. The different aspects of X-ray \citep[or accretion disk, see][]{one09} radiation from SMBs can be considered \citep[see in more details][and references therein]{mch10,gil10}, but here we will shortly discuss the broad Fe K$\alpha$ line emitted from SMBs.

 Particularly, \cite{yl01} discussed possibility to use the shape of the Fe K$\alpha$ line to explore possibility of the binary black hole. They found
that the line profiles of iron line may be very complex, with several peaks. They concluded that the AGNs with unusual Fe K$\alpha$ line profiles, which also have short time X-ray variability (order of a couple of hours) may be a good candidate for SMBs \citep[see in more detail][]{yl01}.
 The spin axes of the two SMBHs are very likely to be misaligned \citep[][]{b80,yl01}, therefore the accretion disks in this system associated with
the SMBs can also have different inclination angles to the line of sight as well as different parameters.
Taking this into account one can expect that SMBs are  expected to emit a very complex Fe K$\alpha$ emission.

\begin{figure}
\centering
\includegraphics[width=2cm,angle=-90]{xradi05.eps}
\includegraphics[width=2cm,angle=-90]{xradi45.eps}
\includegraphics[width=4cm]{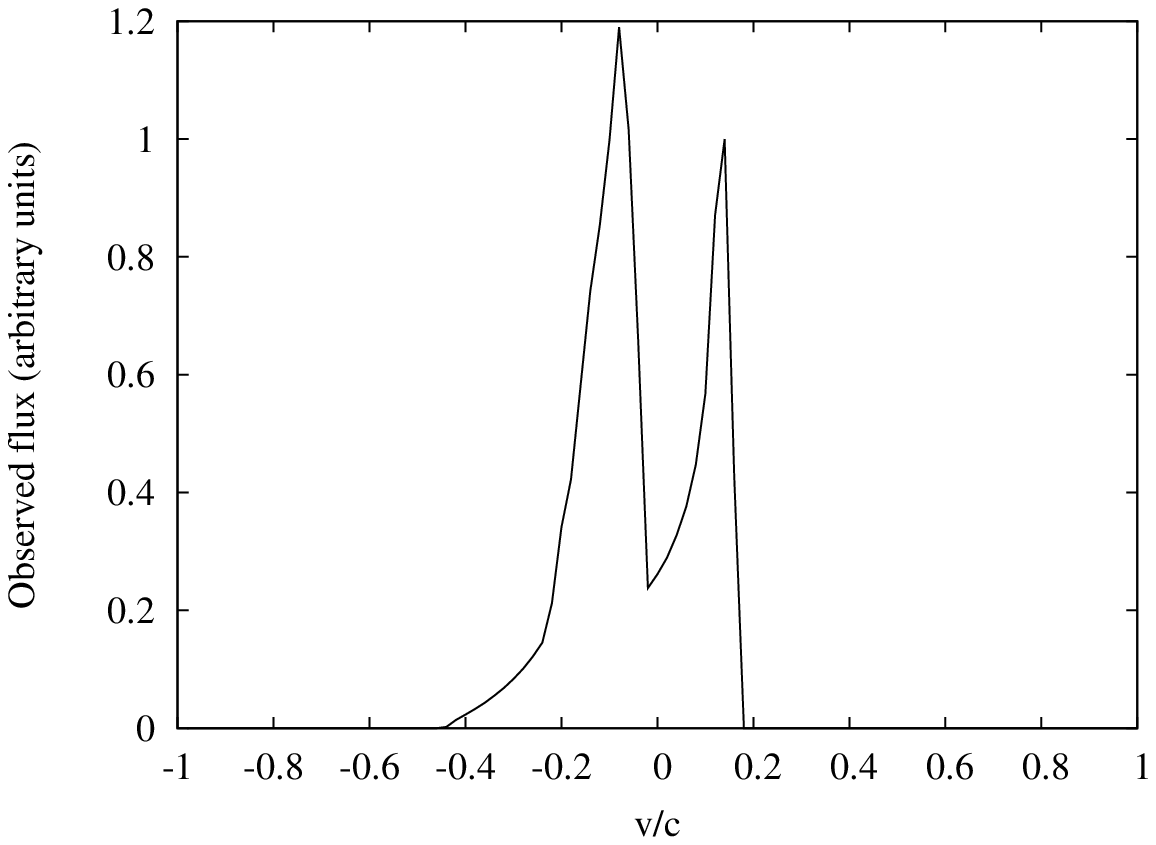}
\includegraphics[width=4cm]{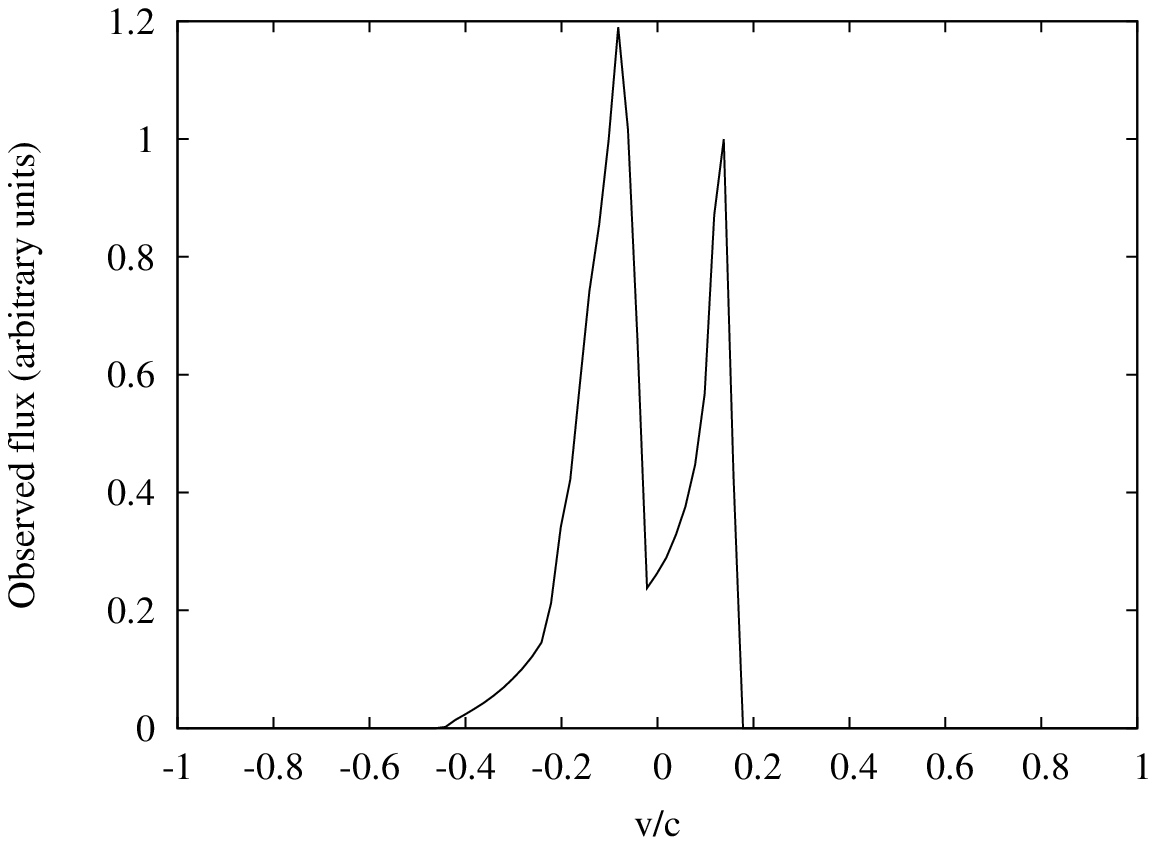}
\includegraphics[width=4cm]{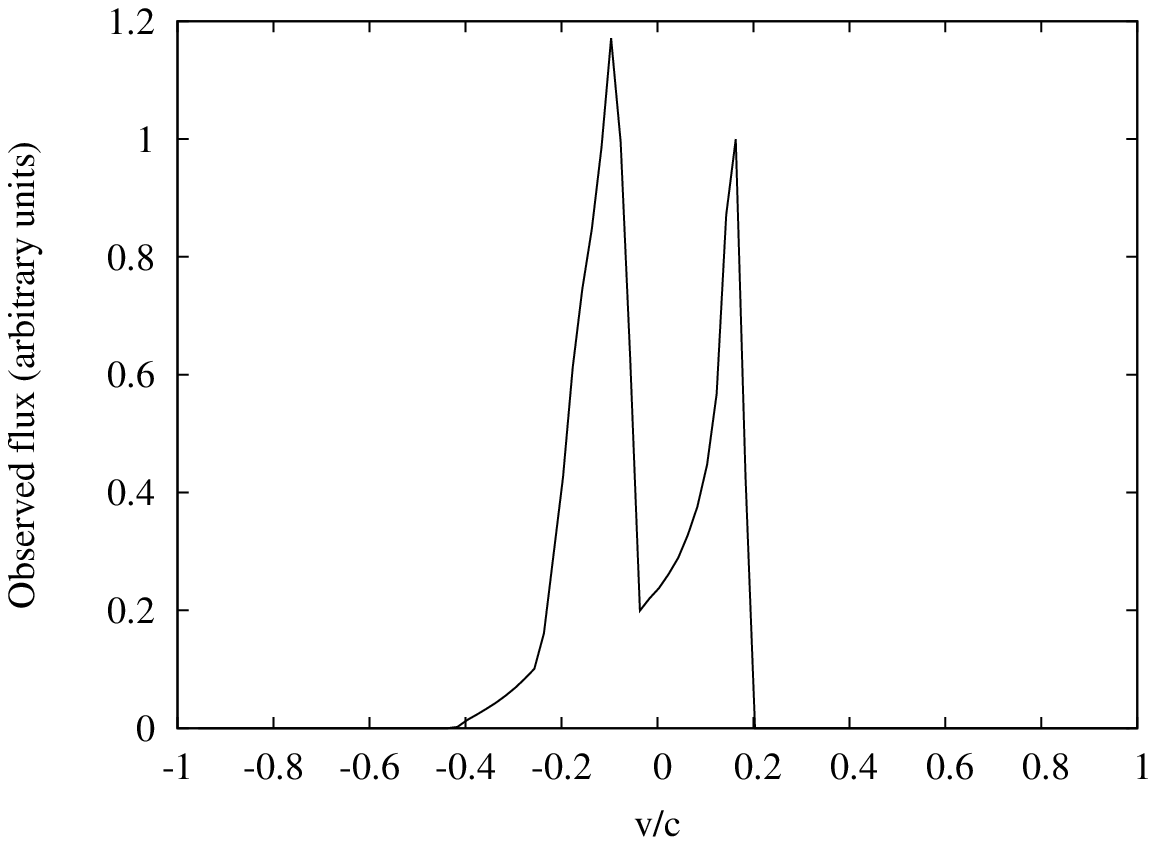}
\includegraphics[width=4cm]{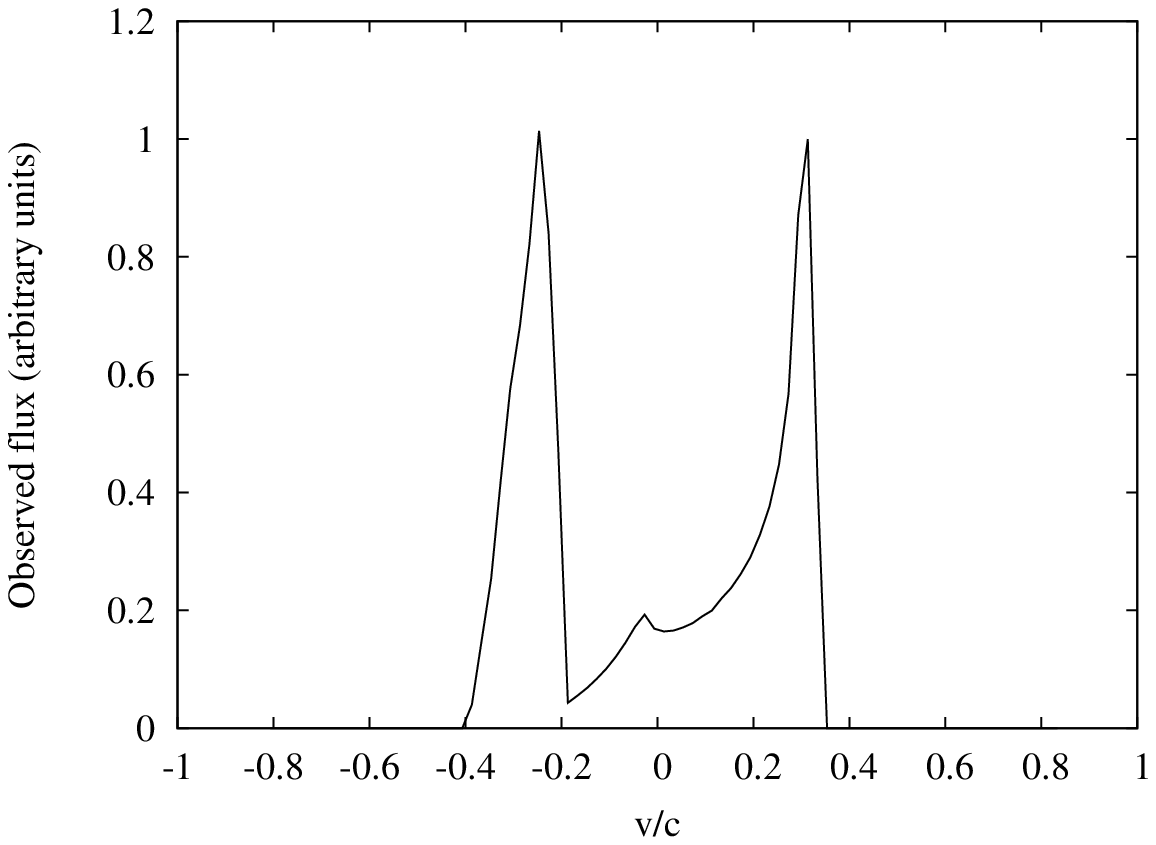}
\caption{Four panels in the first row-up: The simulation of the accretion disks  and corresponding Fe K$\alpha$ line profiles  for the same dimensions of the disk R$_{out}$=50 Rg, but for different inclination of five degree (two panels left up) and 50 degrees (two panels right up). Panels below this: The simulated total line profiles emitted from a SMB system with different distances, i.e., maximal radial velocities 0, 500, 5000 and 50000 km s$^{-1}$, from the left to the right (up to down), respectively.}
\label{f01}
\end{figure}

 Assuming that Fe K$\alpha$ is coming from accretion disks very close to SMBHs we modeled the resulting profiles (see Fig. \ref{f01}). We assumed emission of one near face
inclined disk (inclination 5 degrees) and an additional disk line emitted from a disk with inclination of 45$^0$. We used ray-tracing method \citep[see in more details][and references therein]{p03,pj09}, taking that both disks have the same dimensions of 50 Rg (gravitational radii as outer radius) and that emission is coming from the first stable orbit for a rotating black hole with spin $a$=0.5. We take different distances between two black holes, and consequently different maximal orbital velocities for the two SMBHs with equivalent masses (M$_{BH}=10^8\rm M_\odot$). The accretion disks and emitted line profiles in the rest frame are presented in Fig. \ref{f01}

As one can see in Fig. \ref{f01}, even in the case of small radial velocity (in the case of different accretion disk orientation), the composite Fe K$\alpha$ profile will result in double-peaked profiles, and it seems that Fe K$\alpha$ can be used for detection of SMBs.

\begin{figure}
\centering
\includegraphics[width=7.6cm]{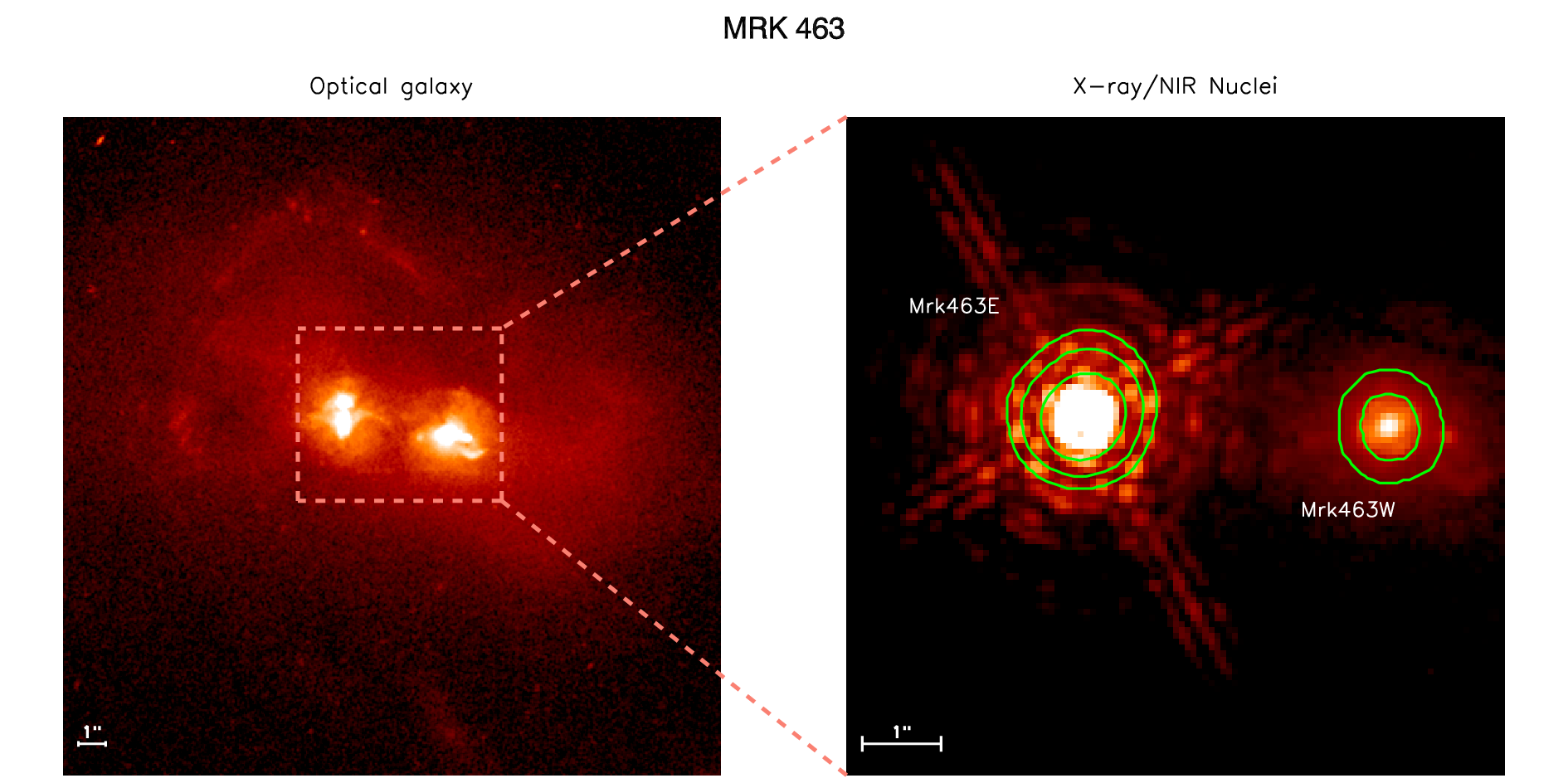}
\includegraphics[width=7.5cm]{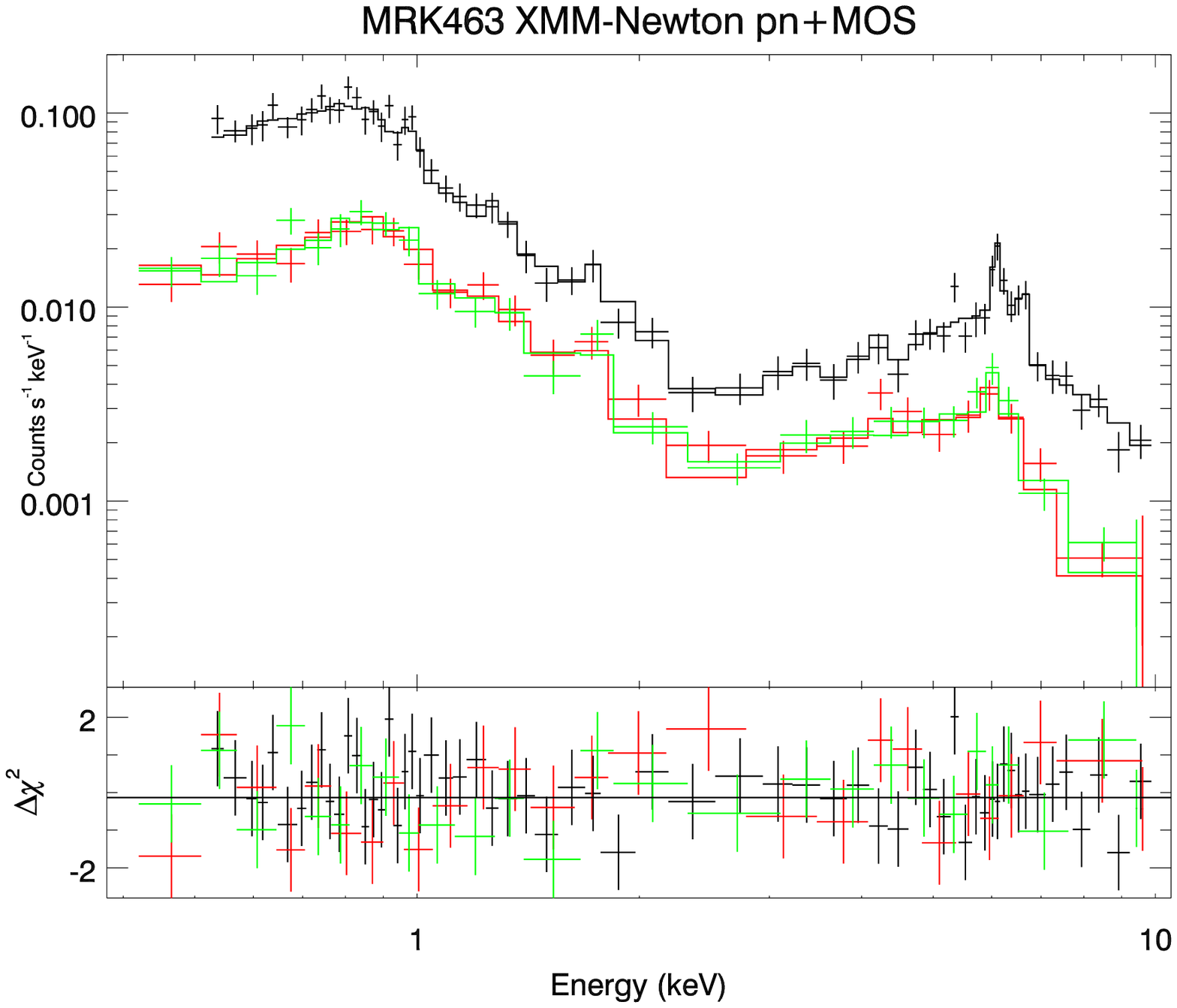}
\caption{Mrk 463 (north is up, east to the left). Up left panel : HST optical (filter f814w) image of this double-nucleus galaxy. Up right panel :
hard X-rays Chandra contours superimposed on the 2.1$\nu$m HST image: two unresolved nuclei are clearly detected in both bands. {\it Down}:.XMM-Newton EPIC, MOS1 and MOS2 spectra of Mrk 463 \citep{b08}}
\label{f11}
\end{figure}

 There are several observations where 'unusual' (shifted or multi-peaked) Fe K$\alpha$ are reported, as e.g., NGC 4151 and NGC 3516, it was found that two
disks with different inclinations better fit the Fe K$\alpha$ line profile \citep[see e.g.][]{w99}. As in the case of NGC 4151, \cite{w99} found that an alternative explanation is a model consisting of a narrow core and two disk lines with inclinations of 58$^0$ and 0 degree, respectively, that also may indicate a SMB system. There are also several other examples: Mkn 766 where the Fe K$\alpha$ line showed variation in the profile \citep[see][]{t07,m07} and PG 1402+261 where line centroid energy at 7.3 keV appears blueshifted with respect to the Fe K$\alpha$ emission band and the blue wing of the line extends to 9 keV in the quasar rest frame \citep[][]{re04}.
As an illustration of the binary X-ray nucleus and corresponding (composed) spectra (with the Fe K$\alpha$ line) is present in Figure \ref{f11}. AGN Mrk 463, has double-nucleus that was resolved with the HST (panel up-left) and Chandra (panel up-right) and composite X-ray spectrum obtained by XMM Newton is present in panel down, where the double-peaked Fe K$\alpha$ profile is present \citep[][]{b08}.

Additionally, there are possible (quasi)oscillation modes in circumbinary disks around eccentric and circular binaries. E.g., \cite{ho09} found that some kind of shock waves can be induced in the inner part of the circumbinary disk by the tidal potential of the binary. These waves may cause variability in emission line profiles emitted from such disks. Moreover a circumbinary disks around SMBs may have different characteristics (structure) than disk around a single SMBH \citep[][]{ls10}, and consequently it can affect the Fe K$\alpha$ line profile. Therefore, (quasi)periodical oscillations (QPOs) may be indicator of SMBs presence \citep[see more detailed discussion in][]{mch10}. Taking rapid variation in the X-ray and huge amount of high-energy radiation \cite{dep03} pointed out three AGNs (Mkn 501, Mkn 421 and Mkn 766) as potential candidates for SMBs, but note here that instabilities in the accretion disk around single SMBH can lead to the quasi-periodical oscillations and very rapid variability.

There are many examples about shifted (or multi-peaked) Fe K$\alpha$, and here is out of scope to discuss all of them, only we mentioned some problems which appear with using the Fe K$\alpha$ to detect SMBs. First of all, only small fraction, around 40\%, of type 1 AGNs emits broad Fe K$\alpha$ line \citep[][]{n97,n07,l10}, second, as a rule the intensity of the broad Fe K$\alpha$ is decreasing with luminosity \citep[]{n97}. Third, the spectral resolution in the X-ray band is too low to detect some fine structures in the line profile, and finally, there is similar problem as in the case of the broad emission lines, the emission may partly come from the accretion disk and partly from an additional region, or even non-axisymmetric illumination can affect the disk line profiles \citep[see][]{yul00}.

\section{Unusual constellation of SMBHs: emission lines of multi-SMBHs, binary quasars and lensed SMBs}

Here we briefly discuss various unusual object (SMB) constellations which can be (probably rarely) present in the nature, but where spectral lines can be used to clarify the nature of objects.

\subsection{Multiple SMBHs}

One possibility is that at pc-scale, as a final merger stage, there are more than two SMBHs \citep[][]{hol07,ama11,kl11}. Emission lines from such, complex SMBH system, would result in a very complex shape, but also, there is a possibility that only one of them are able to emit lines. On the other hand, theoretically  could be possible that, at the same time, recoiling and SMB system may be present (in a multi-SMBH system).

 Triple AGNs on tens kpc \citep[][]{djo07} and  several kpc-scale \citep{li11} were observed, as e.g., \cite{li11} reported the discovery of a kpc-scale
triple AGN, SDSSJ1027+1749 at z = 0.066. The galaxy contains three emission-line nuclei, one with offset of  450 and second with 110 km s$^{-1}$ in velocity from the central one. It is interesting that all three nuclei are classified as obscured AGNs based on optical diagnostic emission line ratios, with black hole mass estimates as around 10$^8 M_\odot$ from stellar velocity dispersions measured in the associated stellar components.

In a multi component SMBH system, one can expect more complex situation, as e.g., line with three peaks, or strongly shifted two components. In some AGNs, as e.g., in SDSS J153636.22+044127.0 three peaks were found \citep[][]{c10}. However up to now, only kpc-scale triple systems have been observed \citep[see][]{mor99,djo07,c10,li11}.

\subsection{Binary quasars, SMBs and gravitational lenses}

 Binary quasars are also the matter of debate; whether they are interacting systems and how much they are connected with SMBs
\citep[][]{koc99,mor99,fvd09}. There are a number of binary quasars \citep[][]{henn06} and they are probably interacting quasars.

\begin{figure}
\centering
\includegraphics[width=8cm,angle=-90]{fig-lens.ps}
\includegraphics[width=8cm]{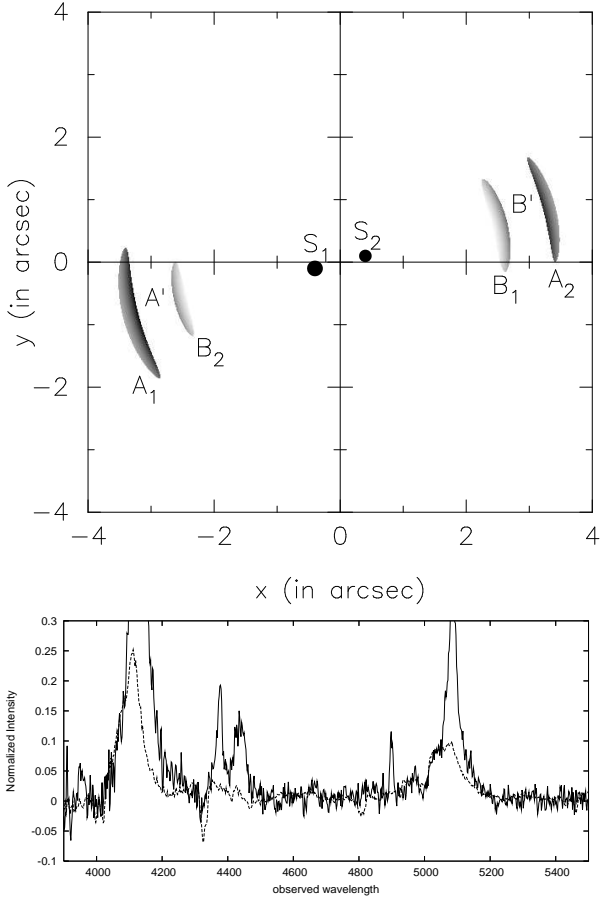}
\caption{Up: Gravitational lens effect on the close sources. Down: The spectra of RXJ 0921+4529, where the lines from the component B (reduced) well fit the line wings of the component A.\citep{pop10}}
\label{f-lens}
\end{figure}

On the other hand, the phenomenon of gravitational lens gives us possibility to observe (and resolve) sources which will be out of sensitivity of our telescopes without amplification caused by the lens effect. This effect may help us to investigate an unresolved BLR structure \citep[microlensing effect in the broad emission lines, see][]{aba02,slus11}. The question here is: Can gravitational lens help us to detect a sub-parsec SMBs? In principle it may be a case in nature.  Hypothetically, there is possibility, as it is shown in Fig. \ref{f-lens} (up), that a SMB is lensed by foreground galaxy, and because of different amplification of SMBHs (let us say SMBH$_{1,2}$) emitting regions, observational effect can be as following: In one component (image) the emission from the region of SMBH$_1$ is dominant and in another image the emission of the SMBH$_2$ emitting region. This may result in different line profiles in two images of a lensed SMB, since there are velocity drift (due to orbital motion) and different line shapes between emitting regions around SMBHs. Consequently, one may misleadingly conclude that there is, e.g., a binary quasar instead of a lensed SMB at pc or sub-pc scale.

 An example is binary (or lensed) quasar RXJ 0921+4529. As it can be seen in Fig. \ref{f-lens} (down), the line wings of the (reduced) line spectrum of
the component B very well fit the wings of lines in the spectrum of the component A \citep[see][]{pop10}. Here may be a possibility that RXJ 0921+4529 is a SMB system with velocity drift around 2700 km s$^{-1}$ (obtained from difference in the redshift $\Delta z\approx0.009$) between two SMBs. Howevlser, such high velocity drift indicates that here is probably a visually close binary \citep[see][]{dr10} or binary quasar.

At the end, the lensed SMB system may, in perspective, be observed, but the problem is to find method to resolve two broad line components, especially if the amplification of two SMBH emitting regions is nearly the same. On the other hand, a lensed  SMB systems may appear as a binary quasar, since the line profiles (and shifts) may be different in two images of a SMB. This effect may be   interesting to investigate in the near future.

\section{Conclusions}

Here we consider emission lines as a tool to detect  SMBs. Mainly we considered the broad emission lines which are emitted from the BLR. The basic problem with broad emission lines, is that the nature (geometry and structure) of BLRs in the, so called, ordinary BLR (where only one SMBH is present) is still purely known. There are several models which can be applied to explain broad emission lines, but we focused here on models which could explain unusual double- (multi-) peaked line profiles and highly shifted broad lines with respect to narrow component.

Considering the unusual broad emission line  profiles we can outline following conclusions:

i) The unusual double-peaked profiles can be explained by a disk (or off-axis) model, taking into account different additional effects, but some specific characteristics as e.g.  H$\alpha$/H$\beta$ flux ratio, variability in the broad line profiles may indicate  some SMB effects. In any case investigations of the objects with unusual double-peaked lines and their understanding is necessary in order to understand the nature of disk emission and its possible connection with SMBs.

ii) The high shifted broad emission lines (asymmetric or nearly symmetric) can also be modeled with specific BLR models \citep[as e.g., off-axis model][]{gas10}, but also there is a possibility that such line profiles indicate SMBs (where the BLR has been associated with one SMBHs) or a kicked-off SMBH.

iii) In both peculiar cases, mentioned above, one should find additional evidence about presence of SMBs that may be found in the other part of spectra, as e.g., in the (quasi-)periodical oscillations

iv) Beside unusual line profiles, the SMBs may emit asymmetric (even symmetric) slightly off centered line profiles similar to the ordinary BLR. This opens a question: is there a larger number of SMBs than AGNs with unusual broad line profiles?

v) Variability in the line profile is expected in the case of SMBs, but it is likely that this variation is present in the so called semi-detached SMB system. It is unlikely to observe variability in the broad line profile caused by orbital motion of the SMB components.

vi) The line intensity ratios emitted from the region (from) around a SMB system seem to be different than ones emitted from a region with single SMBH, and it would be interesting to investigate this effect as a possibility to detect SMBs.

Also, shortly we discuss the narrow emission lines and Fe K$\alpha$, an X-ray, line and we can outline following:

i) The double-peaked narrow emission lines may indicate a kpc-scale SMB, but it seems that the double-peak origin in narrow emission lines is caused by the complex velocity field in the NLR (or superposition of two visually close AGNs). However,  unusual forbidden narrow line ratios and shapes may be present in  a pc, sub-pc SMB system.

ii) The broad Fe K$\alpha$ line may be very perspective  for sub-pc scale SMB detections, but there are some problems, as e.g., spectral resolution, complex disk emission, etc. The (quasi)periodical oscillations in the X-ray spectra (consequently in the Fe K$\alpha$ line) emitted from a SMB system  are expected.

Additionally,  we discussed (probably rare) situations that emission from a SMB is amplified by a gravitational lens. It would be interesting to explore such possibility in known lensed and binary quasars.

At the end, in general, the broad spectral emission lines of AGNs may indicate presence of SMBs, but there are many problems that should be solved, first of all it is the nature of the BLR, that is not purely understand yet. Consequently, all observed  unusual broad line profiles suspected to be from SMB (or recoiling SMBH) candidates may be very well explained with alternative models of the complex BLR around a single SMBH.

\

\

\

{\bf Acknowledgment}

This works is supported by Ministry of Education and Science of R. Serbia thought project 176001 "Astrophysical spectroscopy of extragalactic objects".
The work was presented as an invited talk at special workshop "Spectral lines and super-massive black holes" held on June 10, 2011 as a part of activity within the frame of COST action 0905 "Black holes in a violent universe" and as a part of the 8th Serbian Conference on Spectral Line Shapes in Astrophysics. I would like to thank to Martin Gaskell and Michael Eracleous for a very useful  and critical discussion and suggestions which helped me very much in writting this paper.

\end{document}